\documentclass[aps, prd, amsmath, floats, floatfix, twocolumn,
superscriptaddress, nofootinbib, showpacs]{revtex4-1}

\usepackage[xetex]{graphicx}
\usepackage{color}
\usepackage{soul}
\usepackage{url}
\usepackage{bm}         
\usepackage{times}

\newcommand{\beq}{\begin{equation}}
\newcommand{\eeq}{\end{equation}}
\newcommand{\beqn}{\begin{eqnarray}}
\newcommand{\eeqn}{\end{eqnarray}}

\newcommand{\Caltech}{\affiliation{TAPIR, Walter Burke Institute for Theoretical Physics, MC 350-17,
    California Institute of Technology, Pasadena, California 91125, USA}}
\newcommand{\Einstein}{\affiliation{NASA Einstein Fellow}}
\newcommand{\Cornell}{\affiliation{Center for Radiophysics and Space
    Research, Cornell University, Ithaca, New York, 14853, USA}}

\newcommand{\CITA}{\affiliation{Canadian Institute for Theoretical 
    Astrophysics, University of Toronto, Toronto, Ontario M5S 3H8, Canada}}

\newcommand{\LBL}{\affiliation{Lawrence Berkeley National Laboratory,
1 Cyclotron Rd, Berkeley, CA 94720, USA}}
\newcommand{\NCSU}{\affiliation{Department of Physics, North Carolina State University, Raleigh, North Carolina 27695, USA; Hubble Fellow}}

\begin{document}

\title{Impact of an improved neutrino energy estimate on outflows in neutron star merger simulations}

\author{Francois Foucart}  \LBL \Einstein%
\author{Evan O'Connor} \NCSU 
\author{Luke Roberts} \Caltech \Einstein
\author{Lawrence E. Kidder} \Cornell
\author{Harald P. Pfeiffer} \CITA
\author{Mark A. Scheel} \Caltech

\begin{abstract}
Binary neutron star mergers are promising sources of gravitational waves for ground-based detectors such as Advanced LIGO.
Neutron-rich material ejected by these mergers may also be the main source of r-process elements in the Universe, while radioactive decays in the ejecta
can power bright electromagnetic post-merger signals. 
Neutrino-matter interactions play a critical role 
in the evolution of the composition of the ejected material, which significantly impacts the outcome of nucleosynthesis and the properties of the associated
electromagnetic signal. In this work, we present a simulation of a binary neutron star merger using an improved 
method for estimating the average neutrino energies in our energy-integrated neutrino transport scheme.
These energy estimates are obtained by evolving the neutrino number density in addition to the neutrino energy and flux densities.
We show that significant changes are observed in the composition of the polar ejecta when comparing our new results with earlier simulations
in which the neutrino spectrum was assumed to be the same everywhere in optically thin regions. In particular, we find that material ejected
in the polar regions is less neutron rich than previously estimated. Our new estimates of the composition of the polar ejecta make it more likely that the
color and timescale of the electromagnetic signal depend on the orientation of the binary with respect to an observer's line-of-sight. 
These results also indicate that important observable properties of neutron star mergers are sensitive to the neutrino energy spectrum, and may need to be studied 
through simulations including a more accurate, energy-dependent neutrino  transport scheme.
\end{abstract}


\maketitle

\section{Introduction}
\label{sec:intro}

The detection of gravitational waves from binary black hole mergers
by Advanced LIGO~\cite{Abbott:2016blz,Abbott:2016nmj} just opened a new window through which to
observe the universe. In the coming years, Advanced LIGO~\cite{aLIGO2}
and its European and Japanese counterparts, Advanced VIRGO~\cite{aVirgo2}
and KAGRA~\cite{kagra}, are expected to detect neutron star-neutron star (NSNS) and
neutron star-black hole (NSBH) mergers~\cite{Abadie:2010cfa}. Compact binary mergers in the presence
of at least one neutron star could put strong constraints on the equation of state of nuclear matter
in the extreme conditions reigning in the core of neutron stars~\cite{Read2009b,DelPozzo:13,Lackey2014}.
They are also likely to be the progenitors of short gamma-ray 
bursts~\cite{moch:93,LK:98,Janka1999,Fong2013}, 
and are followed by bright radioactively powered 
optical/infrared~\cite{1976ApJ...210..549L,Li:1998bw,Roberts2011,Kasen:2013xka,Tanaka:2013ana}
and radio transients~\cite{Nakar:2011cw,Hotokezaka:2016} which could provide us with useful information about
the merging objects and their environment. Finally, matter ejected during a neutron star 
merger is a prime candidate for the so-far unknown site of r-process nucleosynthesis, 
where many heavy elements (e.g. gold, uranium) are 
produced~\cite[e.g.][]{korobkin:12,Wanajo2014}.

Numerical simulations are a critical tool to understand the gravitational wave 
and electromagnetic signals powered by NSBH and 
NSNS mergers. Yet, the complexity of the physical processes
playing an important role in these mergers places significant limitations on
the realism of existing simulations. General relativity, magnetohydrodynamics,
neutrino radiation, and nuclear physics all influence at least some important
observables of these systems. Considering the high computational cost of including each
of these components, simulations have generally focused on a subset of these
physical processes, either by improving the microphysics with approximate
treatments of gravity, or using full general relativity with much simpler
physics (see~\cite{Duez:2009yz,faber:12,2014ARA&A..52..661L,BaiottiReview2016} for reviews of numerical simulations). 
Recent general relativistic simulations have only begun 
to partially resolve the effects of magnetic 
fields~\cite{Rezzolla:2011da,Kiuchi2014,Kiuchi2015,2016arXiv160402455R,2015arXiv150202021D}, 
to include
approximate treatments of the neutrinos and better equations of state for dense
matter~\cite{Sekiguchi:2011zd,Neilsen:2014hha,Sekiguchi:2015,Foucart:2015,2016arXiv160300501L,RadiceLeak:2016,Sekiguchi:2016}, 
or both (with sub-grid models for the growth of magnetic 
fields)~\cite{Palenzuela2015}.

In this paper, we focus on the treatment of the neutrinos and their impact on the
post-merger properties of NSNS mergers. Neutrinos are
particularly important as the main source of cooling in the post-merger remnant.
They also play a critical role in setting the relative number of neutrons and protons
in the remnant and in the material ejected from the system. The composition of the fluid is needed to predict
the properties of optical/infrared transients powered by r-process nucleosynthesis
in material ejected by the merger~\cite{Kasen:2013xka,2013ApJ...775...18B}, 
as well as the relative abundances of the
r-process elements produced in the ejecta~\cite{Wanajo2014,Lippuner2015}. 
Finally,
neutrinos can drive strong winds from the post-merger 
remnant~\cite{Dessart2009,Just2014,Perego2014,Sekiguchi:2015,FoucartM1:2016}.

Neutrinos were first included in general relativistic simulations of neutron star mergers 
through a simple
gray (i.e. energy-integrated) leakage scheme~\cite{Sekiguchi:2011zd}, based on approximate methods developed for 
Newtonian simulations~\cite{1997A&A...319..122R,Rosswog:2003rv}. 
A leakage scheme uses the local properties of
the fluid and an estimate of the neutrino optical depth to determine the amount of
energy lost locally to neutrino-matter interactions, and the associated change in the composition of the fluid.
Leakage schemes provide an order-of-magnitude accurate estimate of neutrino cooling
in the post-merger remnant, and have thus been used to capture the first-order effect of
neutrino-matter interactions in general relativistic simulations of compact binary 
mergers~\cite{Sekiguchi:2011zd,Deaton2013,Foucart:2014nda,Neilsen:2014hha,Palenzuela2015,2016arXiv160300501L,RadiceLeak:2016}. 
In most implementations, they do not account for irradiation
of low-density regions by neutrinos emitted from hot, dense regions. This potentially leads
to large errors in 
the composition of the outflows, mostly by underestimating
the number of protons~\cite{FoucartM1:2015,FoucartM1:2016}. Accordingly, the simplest leakage schemes 
are very
inaccurate when attempting to predict the properties of post-merger electromagnetic signals.
More complex leakage schemes have been developed to attempt to include neutrino
absorptions in low-density regions, either by assuming propagation of the neutrinos
along the radial direction~\cite{RadiceLeak:2016}, or through a more expensive global procedure (only used
in Newtonian physics so far) to estimate where neutrinos are transported~\cite{Perego:2016}. 
The latter scheme also includes a discretization of the neutrinos in energy
space.

The only general relativistic simulations going beyond neutrino leakage 
use a moment formalism with an analytic closure to approximate the Boltzmann equation~\cite{1981MNRAS.194..439T,shibata:11}. 
In particular, neutron star merger simulations
have been performed with a gray M1 scheme~\cite{Sekiguchi:2015,FoucartM1:2015,FoucartM1:2016,Sekiguchi:2016}, 
in which the energy density and
flux density of each neutrino species are evolved. 
In NSNS mergers, the use of this moment formalism showed that a wide range of 
compositions, and thus of nucleosynthesis outcomes, exists in the material ejected by the merger~\cite{Wanajo2014}.
Comparisons with leakage schemes for BHNS~\cite{FoucartM1:2015} and NSNS~\cite{FoucartM1:2016} 
mergers clearly show that irradiation of the polar outflows by neutrinos emitted by
the post-merger remnant causes those outflows to be significantly less neutron-rich than predicted
by a leakage scheme which does not account for neutrino absorption. 

The gray M1 scheme is far from
perfect. One obvious limitation is the impact of the analytical closure, which causes
unphysical ``shocks'' in regions in which neutrinos converge.
This occurs in the polar regions of post-merger remnants, putting into question
the accuracy with which we can recover the composition of the polar outflows in those systems.
Another limitation is the lack of information about the energy spectrum of the neutrinos, or even
their average energy. In~\cite{FoucartM1:2015,FoucartM1:2016}, for example, neutrinos in optically thick regions are assumed to be in equilibrium
with the fluid, which is reasonable, but neutrinos in optically thin regions are assumed to follow
everywhere a blackbody spectrum with a temperature determined from the average properties
of the neutrino radiation predicted by the simpler leakage scheme. This neglects potentially important
spatial variations in the neutrino spectrum, deviations from a blackbody spectrum, and the effects
of relativistic beaming on the neutrino energies. These approximations could easily affect our ability to 
predict the composition of the ejected material, as many neutrino-matter cross-sections scale as the
square of the neutrino energy. Additionally, the transport method used 
in~\cite{FoucartM1:2015,FoucartM1:2016} does not guarantee conservation of the total lepton
number. 

Performing a full merger simulation with an energy-dependent transport scheme, even in the 
relatively simple M1 approximation, is too costly with our current code. 
In this paper, we take an alternative route to assess 
the impact of some of the missing information about the neutrino energies. In addition to the
neutrino energy density and flux density, we now evolve the neutrino number density. This does not
provide us with any information about the shape of the neutrino spectrum, but does
provide a local estimate of the average neutrino energy, and accounts for relativistic beaming.
By evolving the neutrino number density, we can also guarantee conservation of the total lepton number.

We consider in particular a low-mass neutron star merger ($1.2M_\odot-1.2M_\odot$) already
studied with our previous M1 and leakage schemes~\cite{FoucartM1:2016} (hereafter Paper I), 
to facilitate comparisons.
We show that relativistic beaming, spatial variations in the average neutrino spectrum, and an improved
treatment of the diffusion rate of the neutrino number density can play a significant role in the composition 
of the ejected material and of the post-merger remnant. 

We organize the paper as follow. The numerical methods and detail of the improved M1 scheme are
provided in Appendix~\ref{sec:form}. The physical system under consideration and initial conditions
are discussed in Sec.~\ref{sec:setup}. The impact of the neutrino scheme on the emitted neutrino
radiation is presented in Sec.~\ref{sec:neutrinos}. Finally, we discuss consequences on the properties
of the ejected material and associated electromagnetic signal in Sec.~\ref{sec:outflows}, 
and summarize our results in Sec.~\ref{sec:conclusions}.

\section{Numerical setup}
\label{sec:setup}

We consider the merger of two neutron stars, each of gravitational mass $M_{\rm NS}=1.2M_\odot$, 
which allows us to probe the lower end of the expected mass distribution function of neutron stars~\cite{Ozel2012,Antoniadis:2016}.
As the main objective of this work is to provide comparisons between different schemes for the treatment
of neutrinos in NSNS mergers, we consider the binary system already evolved in Paper I with both
a leakage scheme and gray M1 transport. The neutron stars are initially non-spinning, and on quasi-circular
orbits (eccentricity $e~\sim 0.01$). Matter within the neutron star is described by the Lattimer-Swesty equation of
state with nuclear incompressibility parameter $K_0=220\,{\rm MeV}$ (LS220~\cite{Lattimer:1991nc}). We use a publicly
available table for the LS220 equation of state~\cite{stellarcollapseeos,OConnor2010}, 
which provides the properties of matter as a function of density, temperature, and composition. For the LS220 equation of state,
a $1.2M_\odot$ neutron star has a radius of $12.8\,{\rm km}$ and a baryon mass $M_b=1.309\,M_\odot$.
Such a radius falls towards the high end of the radii deemed acceptable by existing astrophysical constraints~\cite{2013ApJ...765L...5S}.
Constraint-satisfying initial data is
constructed using the Spells code~\cite{Pfeiffer2003}, as adapted to binary systems in which matter is present~\cite{FoucartEtAl:2008,Tacik:2015tja}.
From the initial conditions, the neutron stars undergo about $2.5$ orbits before coming into contact. For such a low
mass system, the merger results in the formation of a rapidly rotating, massive neutron star below the maximum
baryon mass allowed for uniform rotation at zero temperature ($M_{\rm max}^b=2.83M_\odot$) for the LS220 equation of 
state, as computed using the code of~\cite{cook92,cook94a}.

\begin{figure}
\flushleft
\includegraphics[width=0.90\columnwidth]{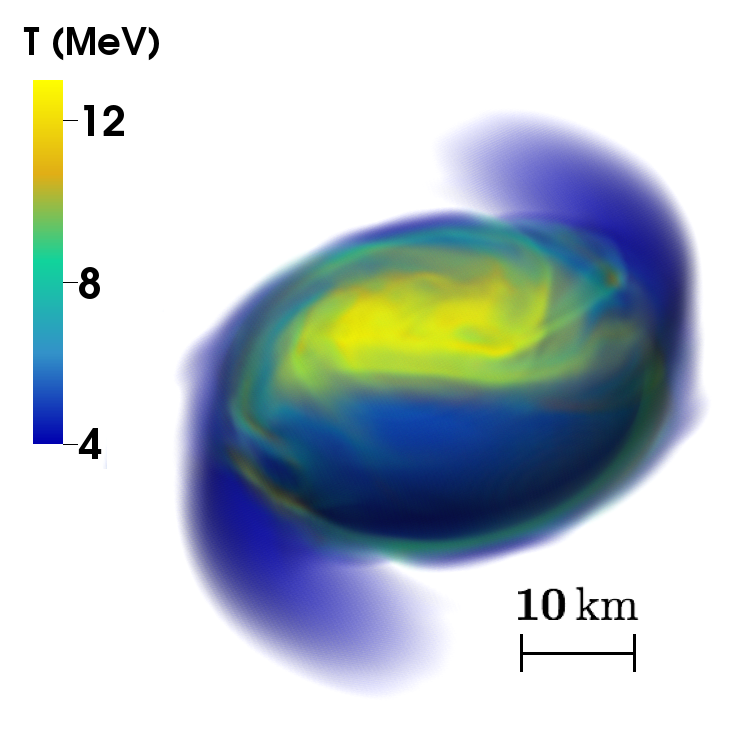}
\caption{3D visualization of the system at the time of merger. The color scale shows the temperature (in MeV),
with denser regions appearing more opaque. Visible regions have a density $\rho \gtrsim 10^{11}{\rm\,g/cm^3}$.
Cold tidal outflows and a small amount of hot shocked material are ejected by the merger, while a stable, hot massive
neutron star forms.}
\label{fig:RhoT-t0}
\end{figure}

\begin{figure}
\flushleft
\includegraphics[width=1.\columnwidth]{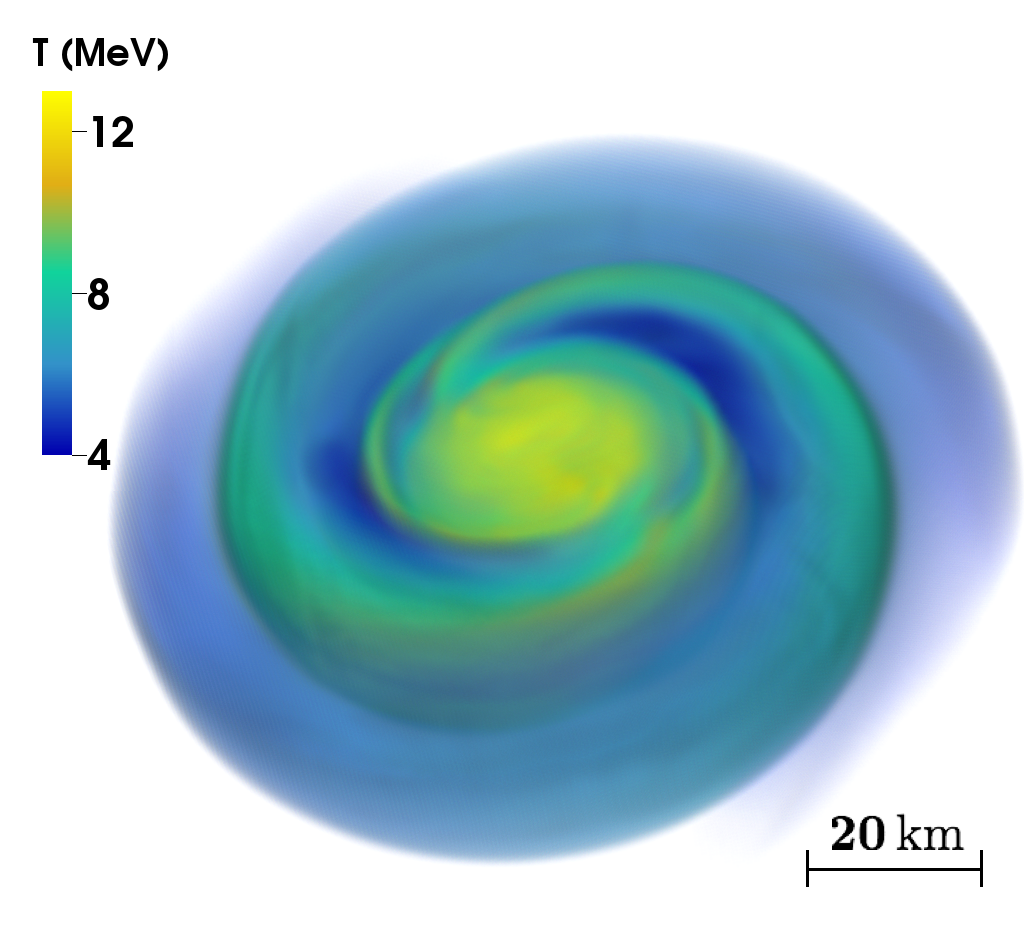}
\caption{Same as Fig.~\ref{fig:RhoT-t0}, but for the remnant $10\,{\rm ms}$ after merger.  
The hot core of the remnant and shocked tidal arms in the disk are clearly visible.}
\label{fig:RhoT-t10}
\end{figure}

We evolve the binary using the SpEC code~\cite{SpECwebsite}, with the same evolution methods as 
in Paper I: a pseudospectral grid 
for the evolution of Einstein's equation in the Generalized Harmonic framework~\cite{Lindblom:2007}, and a finite volume grid
for the evolution of the general relativistic equations of hydrodynamics, written in conservative form~\cite{Duez:2008rb}. The neutrinos
are evolved on the same grid as the fluid. 
Before contact, we use a uniform grid covering all regions of space in which matter is present.
After contact, we use three levels of refinement, with a grid spacing multiplied by a factor of two between each level.
As in Paper I, we use a grid spacing on the finite volume grid of 
$\Delta x_{\rm FV} \sim250\,{\rm m}$ during inspiral and $\Delta x_{\rm FV}  \sim 300\,{\rm m}$ after contact (for the finest level
of refinement).
Each level has $200^2\times100$ cells, taking advantage of the smaller extent of the post-merger remnant in the vertical
direction. We refer the interested reader to Paper I and~\cite{Duez:2008rb,Foucart:2013a,FoucartM1:2015} for more detail about our numerical methods.
The main difference from Paper I is the use of an upgraded M1 scheme, in which the neutrino energies are determined
from the evolution of the neutrino number density (see Appendix~\ref{sec:form}). This allows us to study spatial variations
in the average neutrino energies and the impact of improved energy estimates (including, e.g.,  relativistic beaming and a better estimate
of the diffusion rate of the neutrino number density) on the properties of the material unbound by the merger. 
In all neutrino schemes, we consider 3 species of neutrinos: the electron
neutrinos $\nu_e$, electron antineutrinos $\bar \nu_e$, and a species regrouping the 4 heavy lepton neutrinos
$\nu_x = (\nu_\mu,\bar\nu_\mu,\nu_\tau,\bar\nu_\tau)$, which have similar emissivities and opacities at the temperatures
observed in our simulations.

The main objective of this study is to assess the impact of our new scheme for the evaluation of the neutrino energies
on the main observable of a NSNS merger. Accordingly, we use a numerical setup as close as possible as the one
from Paper I. In Paper I, we performed a lower resolution simulation in order to assess the accuracy of our results.
We found that the properties of the post-merger remnant (composition, temperature, accretion disk profile) 
and the composition and entropy of the ejecta were captured by our simulations with less than $10\%$
relative errors, most likely making numerical errors in those quantities less important than the impact of missing or inaccurate
physics (e.g. exact nuclear equation of state, magnetic fields, neutrino energy spectrum, M1 closure).
However, the mass of cold material ejected in the equatorial plane by the tidal disruption
of the neutron stars is not accurately captured by our simulations, due to the very small amount of material
ejected through that mechanism in equal mass NSNS mergers. 
The same caveat naturally applies to the simulation presented in this work.

We note that, not surprisingly, the overall dynamics of the system is unaffected by the treatment of the neutrinos. This was already
the case when moving from a leakage scheme to the M1 transport scheme, as seen in Paper I,
and remains true here. The main features of the 
merger and post-merger remnant are worth summarizing in order to put into context our discussion of the neutrino 
radiation and matter outflows. A more detailed analysis, as well as comparisons to mergers with different equations of state,
is provided in Paper I. At the time of merger, shown on Fig.~\ref{fig:RhoT-t0}, the contact region between the two neutron stars is rapidly heated, while strong 
tidal arms form in the region of each neutron star antipodal to the contact region. 
These tidal arms contain a small amount of unbound material (the 
exact mass is unresolved in our simulations, but less than $10^{-3}M_\odot$), and a larger amount of bound material.

Within a few milliseconds, the two neutron star cores merge into a distorted massive remnant, with strong excitation
of the fundamental quadrupolar mode of the remnant neutron star. This mode causes the emission of large amplitude gravitational waves, which
if measured can provide tight measurements of the neutron star equation of state~\cite{Bauswein:2014qla,Takami:2015,Clark:2015zxa}. 
Around the same time, the bound
matter in the tidal arms forms a thick, dense accretion disk. The post-merger remnant and accretion disk at the end of the simulation,
$10\,{\rm ms}$ after
merger, are shown on Fig.~\ref{fig:RhoT-t10}. Within the disk, strong $l=2$ perturbations are driven by the
excited massive neutron star. These two spiral arms are hotter than the rest of the disk, with 
$T_{\rm spiral}\sim 9\,{\rm MeV}$ and $T_{\rm disk}\sim 5\,{\rm MeV}$. The spiral arms also show sharp density jumps, with the
density inside the arms being about three times the density outside the arms. The massive remnant, which
was heated at the time of merger, is even hotter with $T_{\rm core}\sim (15-20)\,{\rm MeV}$. The spiral arms and hot neutron star
are the main sources of neutrinos, as discussed in Sec.~\ref{sec:neutrinos}. Over the $10\,{\rm ms}$ of post-merger evolution
performed here, more material is ejected from the outer disk in the equatorial plane, while neutrino absorption
drives a wind in the polar regions. These outflows are discussed in Sec.~\ref{sec:outflows}. The 
measured properties of the emitted neutrinos and of the unbound matter are the main observables which change with our treatment
of the neutrinos, and are the focus of this work. 

\section{Neutrino radiation}
\label{sec:neutrinos}

\subsection{General properties}

\begin{figure}
\flushleft
\includegraphics[width=1.\columnwidth]{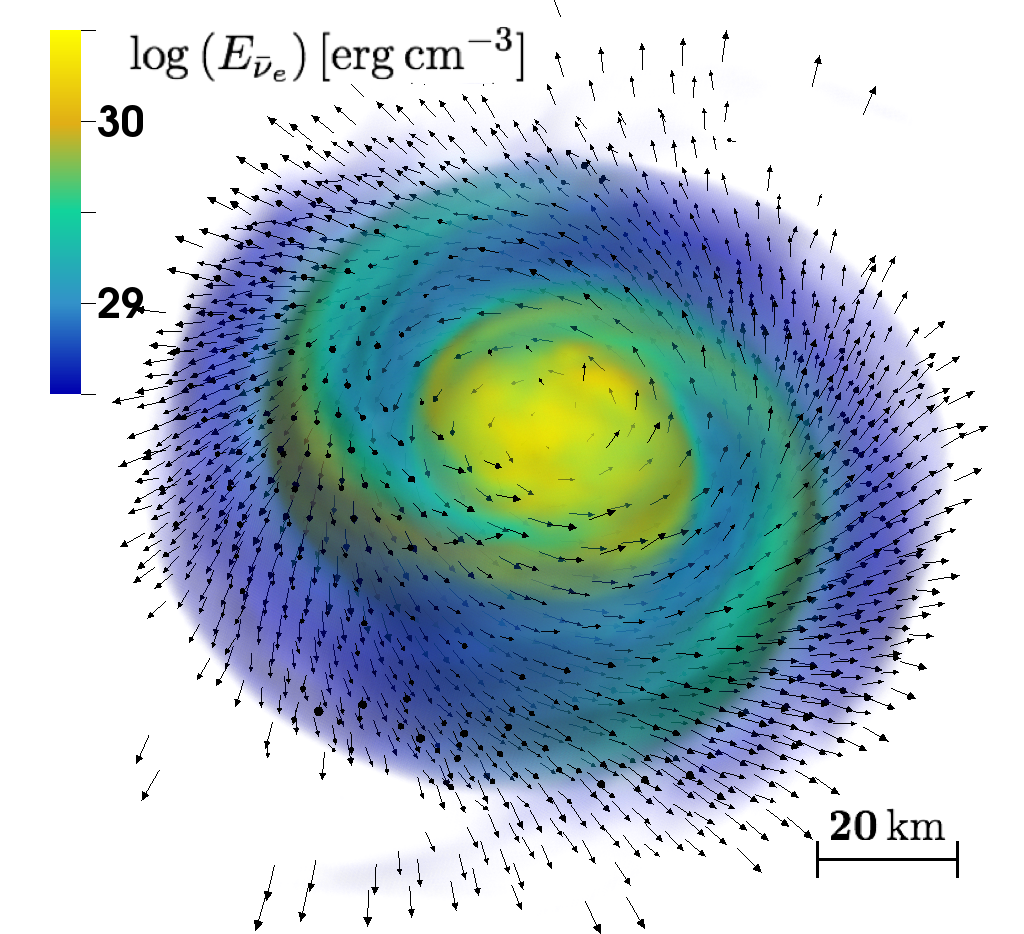}
\caption{Energy density and normalized flux ($\alpha F^i/E - \beta^i$, i.e. the effective transport velocity of the neutrino energy density) 
of electron antineutrinos $\bar\nu_e$ in the high-density regions of the remnant ($\rho\gtrsim 10^{11}{\rm\,g/cm^3}$),
shown $5\,{\rm ms}$ after merger. Most of the neutrino emission comes from the hot core and shocked tidal arms. In the inner disk, 
$\bar\nu_e$ are trapped and advected with the fluid. In the outer disk, they are free-streaming away from the remnant.}
\label{fig:ENuaT5}
\end{figure}

\begin{figure}
\flushleft
\includegraphics[width=1.\columnwidth]{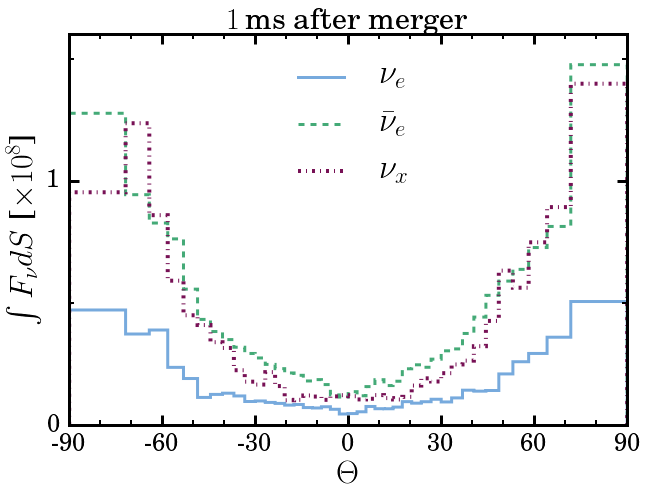}
\caption{Energy flux of neutrinos leaving the computational domain as a function of the angle $\Theta$ between the neutrino flux and the equatorial plane (in degrees), 
$1\,{\rm ms}$ after merger. Results are binned so that each bin represents the same surface area on the unit sphere. The energy fluxes are in units in which $G=c=M_\odot=1$.}
\label{fig:NuTheta1}
\end{figure}

\begin{figure}
\flushleft
\includegraphics[width=1.\columnwidth]{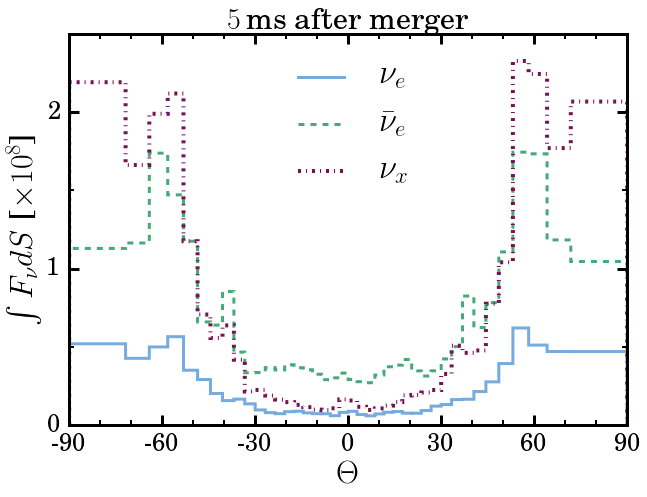}
\caption{Same as Fig.~\ref{fig:NuTheta1}, but $5\,{\rm ms}$ after merger.}
\label{fig:NuTheta5}
\end{figure}

\begin{figure}
\flushleft
\includegraphics[width=1.\columnwidth]{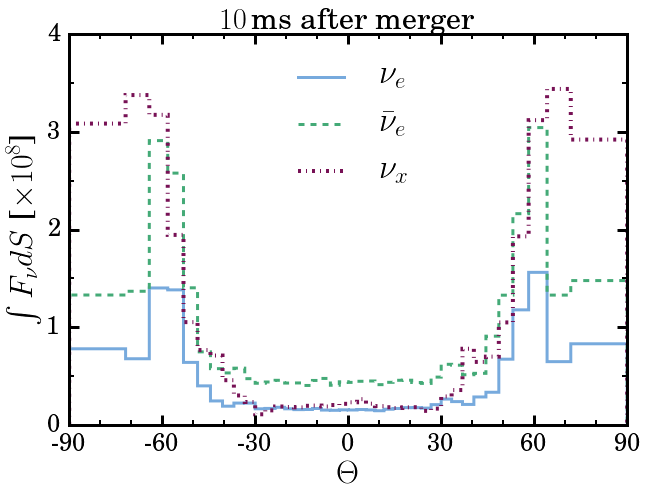}
\caption{Same as Fig.~\ref{fig:NuTheta1}, but $10\,{\rm ms}$ after merger.}
\label{fig:NuTheta10}
\end{figure}

Many of the qualitative properties of the neutrino radiation are independent of
our chosen approximation for the neutrino energy spectrum. In all approximations, the main emission regions are the hot, dense parts of the remnant: the
central core, and the shocked tidal arms. The energy density of $\bar\nu_e$, for example, is shown on Fig.~\ref{fig:ENuaT5}
towards the middle of our post-merger evolution ($5\,{\rm ms}$ after merger). Electron antineutrinos are trapped and advected with the flow
in regions inside the shocked tidal arms. Free-streaming neutrinos in the outer disk are mostly produced in those arms, while
neutrinos in the polar regions come from both the core and the tidal arms. Figs.~\ref{fig:NuTheta1}-\ref{fig:NuTheta10} show the neutrino flux density
as a function of its angle with respect to the equatorial plane $1\,{\rm ms}$, $5\,{\rm ms}$, and $10\,{\rm ms}$ after merger. From these figures,
we can clearly see that most of the neutrinos are initially emitted in the polar directions. 
Once a disk forms, the neutrinos are mostly confined within a cone of $40^\circ$ around the poles, 
with an amplitude peak $30^\circ-40^\circ$ from the poles becoming more visible at later times. This peak is probably due to neutrinos beamed from the shocked
tidal arms, which become less optically thick as time passes. 
The confinement of the neutrinos to the polar directions comes from the fact that neutrinos escape through the low-density 
regions above and below the disk and are confined by the optically thick accretion disk. The exact angular distribution may however be affected by known issues
with the M1 closure when radiation converges from different directions, and should be taken with some caution.

The general properties of the neutrino radiation for $\nu_e$ and $\nu_x$ is similar to what is observed for
$\bar{\nu_e}$.
The fluid is generally more opaque to $\nu_e$ than $\bar\nu_e$, as the disk is very neutron rich. The emission of $\nu_e$ in the equatorial
plane is strongly suppressed, largely due to the fact that the shocked tidal arms are hidden behind material optically thick to $\nu_e$. The polar
luminosity is also about a factor of $2$ lower than for $\bar\nu_e$. The heavy lepton neutrinos $\nu_x$, on the other hand, are nearly
free-streaming as soon as they leave the dense core of the remnant. Most $\nu_x$ are emitted from that dense core, and thermally decouple from the
matter in hotter regions than the $\nu_e$ and $\bar\nu_e$ (see also Fig.~\ref{fig:VertProf}).
As for $\bar\nu_e$, Figs.~\ref{fig:NuTheta5}-\ref{fig:NuTheta10} show that after disk formation most of the $\nu_e$ and $\nu_x$ emission is confined to a cone of about $40^\circ$ in the polar regions. Although beamed emission at a $30^\circ$ angle from the poles still appears to be present, it is not as prominent as for $\bar\nu_e$. This is in keeping with the expectation that a larger fraction of the emitted $\nu_e$ and $\nu_x$ neutrinos come from the dense core.

So far, these results are very similar to what we already observed in~\cite{FoucartM1:2016}, or even qualitatively comparable to the emission regions
predicted by simpler leakage schemes~\cite{FoucartM1:2016,Palenzuela2015}. 
The dynamics of the merger remnant and emission regions of the neutrinos appear to be
robust predictions of both leakage and existing approximate transport simulations. Differences begin to arise when considering the predicted average
neutrino energies, which
we discuss in the next section, and the properties of the outflows, outlined in Sec.~\ref{sec:outflows}. 

\begin{figure}
\flushleft
\includegraphics[width=1.\columnwidth]{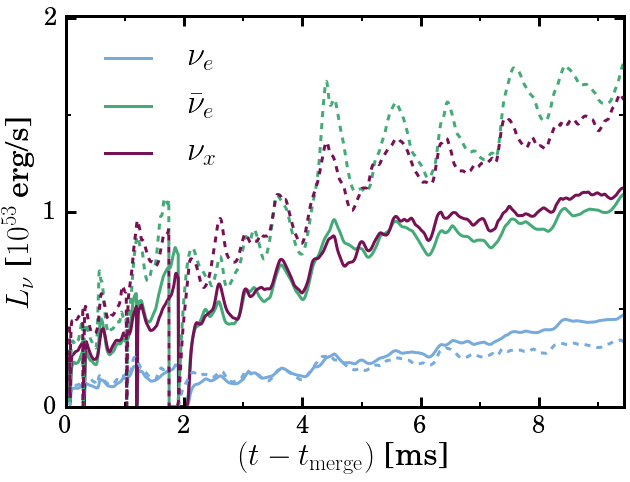}
\caption{Neutrino luminosity, measured as the total energy of neutrinos leaving the computational domain, for the 3 species of neutrinos.
Solid lines show the results with the spatially varying average neutrino energy presented here, while the dashed lines show results with a single global neutrino
temperature in optically thin regions, from Paper I. Emission of $\bar\nu_e$ and $\nu_x$ is significantly
decreased when using a local estimate of the average neutrino energy. We note that here and in subsequent figures, global neutrino quantities are discontinuous 
at $t\sim 2\,{\rm ms}$. This is due to the addition of a lower level of refinement as the matter expand, which leads us to compute the flux of neutrinos out of the grid
on a surface farther from the remnant. Neutrinos take a finite time to propagate from the old measurement surface to the new measurement surface.}
\label{fig:NuLum}
\end{figure}

We also observe that, with our local prescription
for the computation of the average neutrino energy, the luminosity of  $\nu_x$ is decreased by $\sim 30\%-40\%$ compared to the simulation using
a global prescription presented in Paper I (see Fig.~\ref{fig:NuLum}). This is most likely due to a higher estimate of the average 
neutrino energy (and thus higher opacity of the fluid to neutrinos) in this work, as discussed below. 
The luminosity of $\nu_e$ and $\bar\nu_e$ is initially suppressed by $\sim 20\%-30\%$, for the same reason.
At later times, we will see that our current estimates of the average neutrino energy for $\nu_e$ and $\bar\nu_e$ agree better with the results of Paper I.
Yet, at the end of the simulation, the $\bar\nu_e$ luminosity is only $\sim 60\%$  of its value in Paper I. The $\nu_e$ luminosity, 
on the other hand, rises to $\sim 140\%$ of its old value.
This is most likely due to a difference in the evolution of the composition of the remnant, related to a better treatment of the
neutrino number density and, consequently, of the conservation of the total lepton number.

\begin{figure}
\flushleft
\includegraphics[width=1.0\columnwidth]{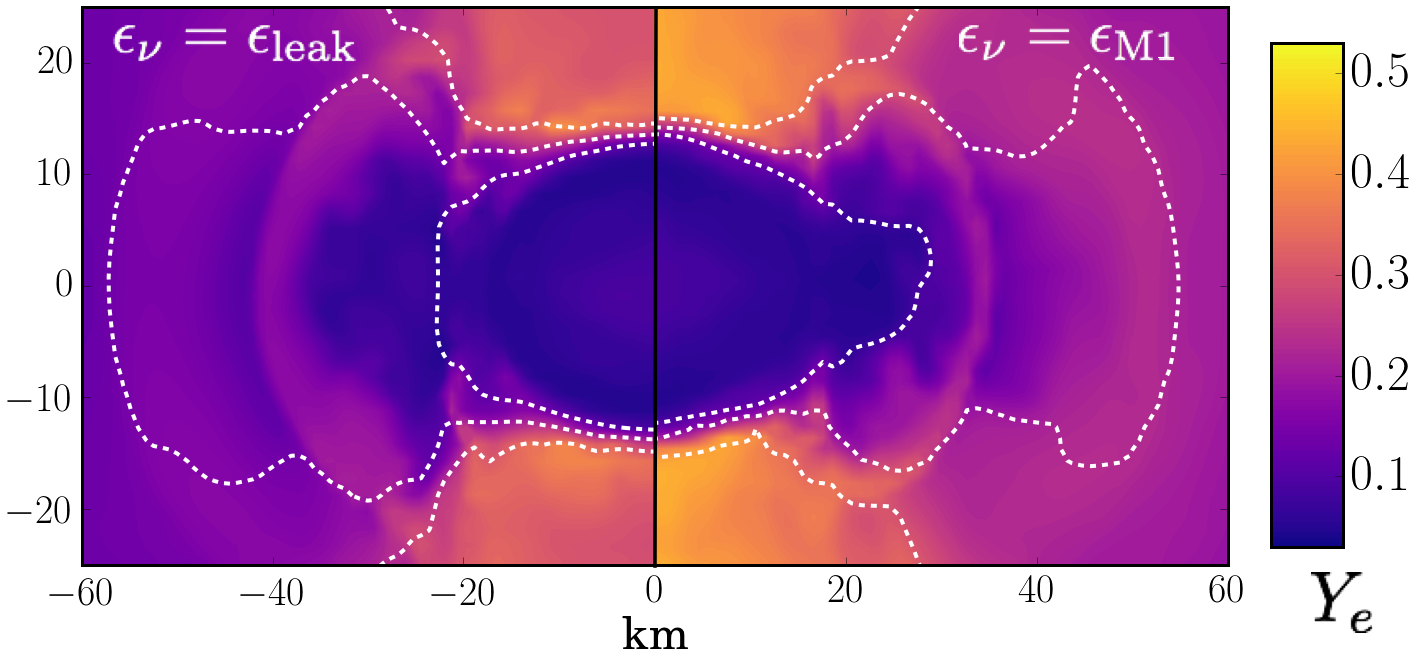}
\caption{Electron fraction of the post-merger remnant $10\,{\rm ms}$ after merger. {\it Left}: Simulation from Paper I, using a global estimate of the neutrino average
energy. {\it Right}: Simulation using a local estimate of the neutrino average energy and evolving the neutrino number density. Dashed lines show density contours
of $\rho = 10^{11},10^{12},10^{13}\,{\rm g/cm^3}$. The latter simulation evolved towards higher electron fractions everywhere but in the core of the 
post-merger remnant. This is a generic feature in our simulations from $\sim 5\,{\rm ms}$ after the merger, with the two simulations slowly diverging over time.
The different density profiles are largely due to minor variations in the phase of the excited mode of the neutron star remnant.}
\label{fig:YeDiff}
\end{figure}

There are a few important effects modifying the evolution of the fluid composition with respect to Paper I. The first is simply the change in our estimate of the neutrino average energies. 
As we will see in the next section, polar regions see higher neutrino energies when the spatial dependence of the neutrinos is taken into account, and will thus absorb 
neutrinos more rapidly. The second comes from the fact that our new transport scheme considers different spectral shapes for the neutrino energy density and the
neutrino flux density, taking into account the faster diffusion of low-energy neutrinos (see Appendix). 
The diffusion of the neutrino number density is better modeled in our new 
scheme, and the composition of optically thick regions will evolve faster than in Paper I. Finally, the simulation presented here consistently evolves the neutrino
number density on the grid. Conservation of the total lepton number is thus guaranteed. The resulting difference in the evolution of the electron fraction of the fluid is
shown on Fig.~\ref{fig:YeDiff}. Except in the core of the post-merger remnant, the fluid evolves towards a higher electron fraction when evolving the neutrino number 
density. This will naturally lead to a relative decrease in $\bar\nu_e$ emission and an increase in $\nu_e$ emission. 

\begin{figure}
\flushleft
\includegraphics[width=1.0\columnwidth]{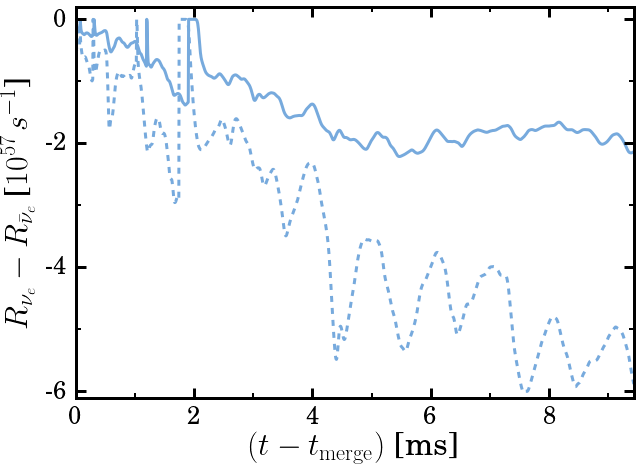}
\caption{Difference between the number flux of $\nu_e$ and $\bar\nu_e$ measured when evolving the neutrino number density (solid line). We also show
the same quantity, but obtained from the neutrino luminosity and estimated neutrino energy used in Paper I (dashed line).}
\label{fig:Nflux}
\end{figure}

The inconsistency in the treatment of the total
lepton number in the simulation from Paper I also leads to unreliable predictions for the number flux of neutrinos leaving the computational domain, as we show in Fig.~\ref{fig:Nflux}. In theory, we expect that the change in the total number of protons on the numerical grid satisfies
 $dN_p/dt \sim - (R_{\nu_e} - R_{\bar\nu_e})$, as both the change in the total number of $\nu_e$ and $\bar\nu_e$ on the numerical grid and the change in $N_p$
 due to mass outflows are small.
We see that Paper I predicted a larger rate of increase of the lepton number within the grid than our simulation evolving the neutrino number density
(which exactly conserves the total lepton number).  
During the last $5\,{\rm ms}$ of
evolution, the change in proton number measured on the numerical grid in the simulation evolving the neutrino number density
is $dN_p/dt \sim 2.1 \times 10^{57}{\rm s}^{-1}$,
which is roughly consistent with Fig.~\ref{fig:Nflux}. In Paper I, the change in proton number on the grid was $dN_p/dt \sim 5.8\times 10^{56}{\rm s}^{-1}$, or ony $10\%
$ of the value estimated in Fig.~\ref{fig:Nflux}. This leads to a lower $Y_e$ in the simulation from Paper I (Fig.~\ref{fig:YeDiff}), despite the neutrino
fluxes indicating stronger emission of electron antineutrinos in that simulation (Fig.~\ref{fig:Nflux}). 
Due to this effect, the compositions of the post-merger remnants in the two simulations slowly 
diverge, starting $\sim 5\,{\rm ms}$ after merger.
The improvement in the conservation of the total lepton number is one of the main advantage of our new transport scheme.

\subsection{Estimated average neutrino energies}

\begin{figure}
\flushleft
\includegraphics[width=1.\columnwidth]{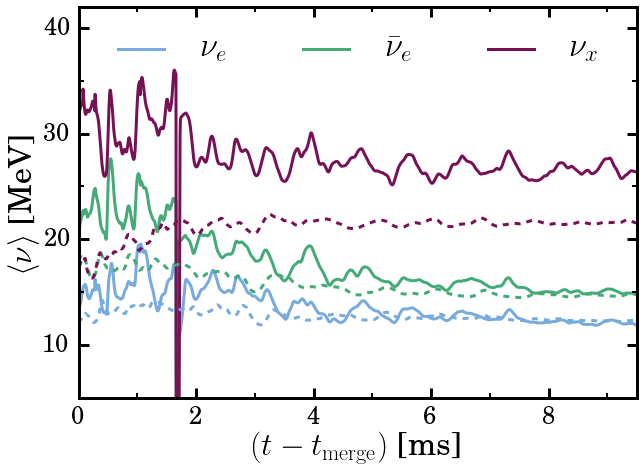}
\caption{Average energy of the neutrinos leaving the computational domain as a function of time for the 3 species of neutrinos. Solid lines show the results with the spatially varying average neutrino energy, while the dashed lines show estimates from the leakage scheme when using a single global neutrino
temperature in optically thin regions, from Paper I.}
\label{fig:NuEavg}
\end{figure}

An important difference between this work and Paper I is the computation of the average neutrino energy. We changed our
estimate for the average neutrino energy from a global estimate based on the predictions of the leakage scheme to a local estimate based on the  evolution of the
neutrino number density. Not surprisingly, this leads to different estimates for the average neutrino energy $\langle \nu \rangle$ of each neutrino species, 
shown in Fig.~\ref{fig:NuEavg}. We see that the local scheme predicts significantly higher average neutrino energies during the merger for all species of neutrinos.
After the formation of a massive disk around the neutron star remnant, the predictions of the leakage scheme agree well with our results evolving the neutrino
number density for $\nu_e$ and $\bar\nu_e$. The heavy lepton neutrinos, however, remain significantly hotter when using the local average energy estimate.

At the time of merger, most of the neutrino emission comes from hot material shocked by the collision of the two neutron stars. That material is moving at a
significant fraction of the speed of light, and away from the contact region. This results in the observed preferential emission of the neutrino emission along the poles
(see Fig.~\ref{fig:NuTheta1}). For an inertial observer in the direction of motion of the emitting fluid (or, if the fluid is optically thick, of the neutrinosphere),
we also expect a Doppler shift between the observed neutrino energy $\langle \nu \rangle$ and the neutrino energy in the emitting fluid element's frame
$\langle \nu_f \rangle$
\beq
\langle \nu \rangle = \sqrt{\frac{1+v/c}{1-v/c}} \langle \nu_f \rangle.
\eeq
The largest shift observed for the heavy lepton neutrinos (see e.g. Fig.~\ref{fig:NuEavg}) 
can easily be explained if the neutrinos are emitted by fluid elements moving at $\sim 0.4c$.

As the accretion disk around the massive neutron star settles down, the velocity Doppler shift begins to play a less important role. Neutrinos coming from the shocked
tidal arms, which are moving at $v \sim 0.3c$, may still be beamed. But the larger fraction of neutrinos coming from the core and emitted along the polar regions
only appear to be subject to a significant Doppler shift at the high densities at which the heavy-lepton neutrinos decouple from the fluid. 

The leakage and M1 schemes are also likely to have different estimates for the the temperature of the region in which the neutrinos thermally decouple from the fluid, particularly for the heavy-lepton neutrinos which thermally decouple long before the surface of last scattering (their absorption optical depth is about 3 orders of magnitude smaller than their scattering optical depth). For $\nu_e$ and $\bar \nu_e$, spatial variations in the temperature of the neutrinosphere as well
as inaccuracies in the approximate leakage scheme can also result in an error in the determination of the temperature of the neutrinosphere, even though the absorption and scattering neutrinosphere are very close to each other.
This error in the temperature of the neutrinosphere can have two important consequences. The first is a change in the predicted average neutrino energies.
The second is an inconsistency in the computation of the neutrino opacities in the scheme used in Paper I. 

\begin{figure}
\flushleft
\includegraphics[width=1.\columnwidth]{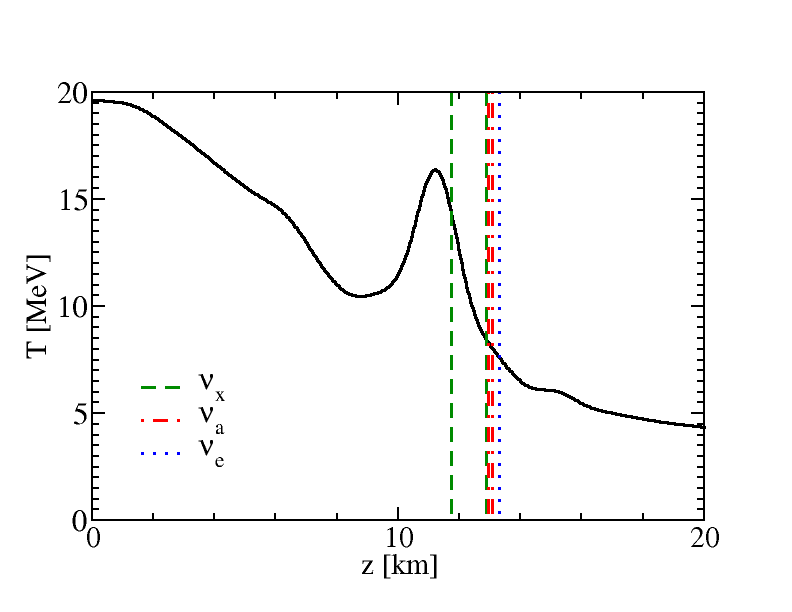}
\caption{Linear profile of the temperature $10\,{\rm ms}$ after merger, along the vertical axis passing through the center of the neutron star.
Vertical lines show the location of the absorption (leftmost line) and scattering (rightmost line) neutrinospheres
for $\nu_x$ (green dashed lines), $\nu_a=\bar \nu_e$ (red dot-dashed lines), and $\nu_e$ (blue dotted line, with both surfaces being undistinguishable).
We note that in Paper I, we assumed that the neutrinos decoupled from the fluid at $T\sim (4,5,7)\,{\rm MeV}$ for $(\nu_e,\bar \nu_e.\nu_x)$, because neutrino
temperature were computed from a global average of the neutrino energies. This is very far from the local value of the absorption neutrinosphere temperature 
obtained in this paper.
For the heavy-lepton neutrinos, we also note a significant difference between the location of the absorption neutrinosphere and the scattering neutrinosphere,
which is a regime in which the grey moment scheme is potentially problematic (see Appendix).}
\label{fig:VertProf}
\end{figure}

If, as observed here, the leakage scheme underestimates the neutrino temperature, then
the transport scheme used in Paper I will underestimate the opacity of the fluid to neutrinos in a region immediately outside of the absorption neutrinosphere. 
Close to the neutrinosphere, the temperature of the fluid decreases with density and the transport scheme from Paper I assumes that
neutrinos thermalize with the fluid as long as the temperature of the fluid is higher than the neutrino temperature {\it predicted by the leakage scheme}. In
Paper I, neutrinos are thus assumed to be thermalized in regions in which the transport scheme would predict that they are already 
thermally decoupled from the fluid, and
hotter than the fluid.
This leads to the following systematic errors in the scheme used in Paper I:  
(1) underestimating the predicted opacity of the fluid; (2) overestimating the neutrino luminosity, due to the smaller optical
depth of the fluid; and (3) overestimating the neutrino number flux, due to the overestimated luminosity and underestimated energy.

To illustrate this point, we show on Fig.~\ref{fig:VertProf} a linear profile of the fluid temperature at the end of our simulation, along the vertical axis
passing through the center of the neutron star (i.e. in a direction in which we have a rapid transition between the high-density neutron star core and
a low-density neutrino-driven wind). 
We also show the location of the absorption and scattering neutrinospheres, estimated from direct integration of the opacities along that vertical axis. 
We note that with the method
used in Paper I, we would have assumed that the neutrinos thermally decouple from the fluid at temperature $T\sim (4,5,7)\,{\rm MeV}$ for 
$(\nu_e,\bar\nu_e.\nu_x)$. We see that along this vertical axis, the temperature of the absorption neutrinosphere was widely underestimated in Paper I.
Fig.~\ref{fig:VertProf} also shows that, in the least favorable direction in which sharp density and temperature gradients are present 
(see e.g. Fig.~\ref{fig:YeDiff} for the density gradient), significant errors in the determination of the neutrino energies are likely: the fluid temperature
varies by $\sim 5\%-15\%$ over a single grid spacing ($\sim 300\,{\rm m}$), and similar errors in the neutrino energies should be expected.

The error in the estimated temperature of the absorption neutrinosphere explains the higher neutrino luminosities observed in Paper I for all 
neutrinos at early times, and for heavy-lepton neutrinos at all times.
The late time decrease in $\bar\nu_e$ emission and increase in $\nu_e$ emission cannot, on the other hand, be attributed to differences in the estimated neutrino
temperature. In the previous section, we showed that they are instead due to differences in the composition of the fluid.

\begin{figure}
\flushleft
\includegraphics[width=1.\columnwidth]{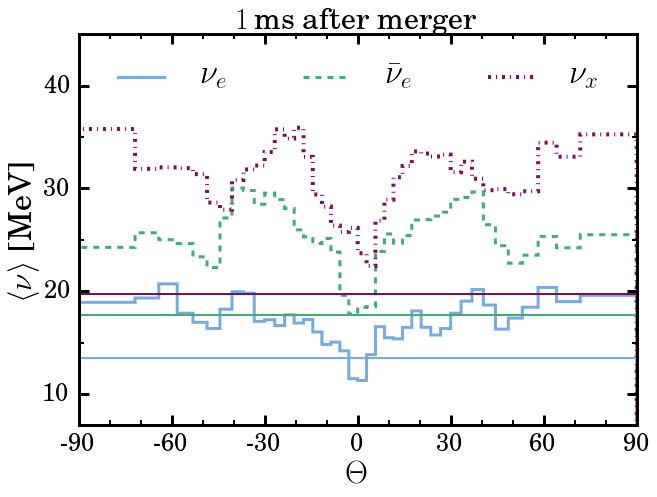}
\caption{Average energy of the neutrinos leaving the computational domain as a function of the angle $\Theta$ between the neutrino flux and the equatorial plane
(in degrees), $1\,{\rm ms}$ after merger. Results are binned so that each bin represents the same surface area on the unit sphere. Solid horizontal
lines show the prediction from the leakage scheme.}
\label{fig:NuEavg1}
\end{figure}

\begin{figure}
\flushleft
\includegraphics[width=1.\columnwidth]{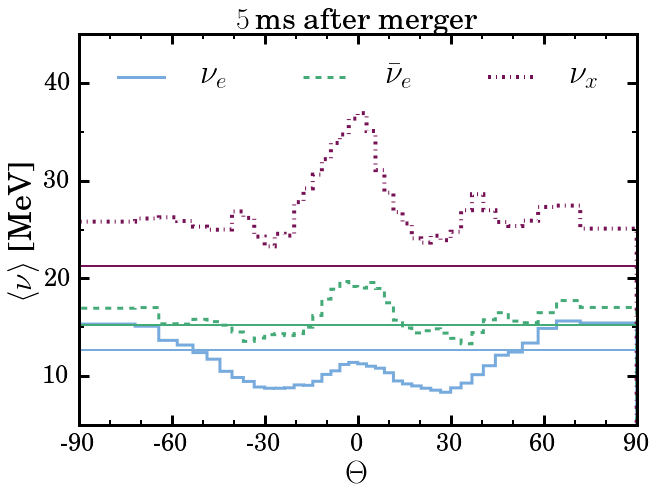}
\caption{Same as Fig.~\ref{fig:NuEavg1}, but $5\,{\rm ms}$ after merger.}
\label{fig:NuEavg5}
\end{figure}

\begin{figure}
\flushleft
\includegraphics[width=1.\columnwidth]{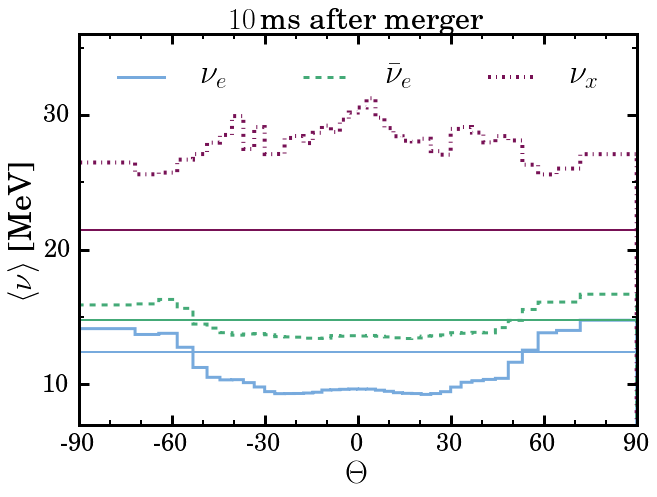}
\caption{Same as Fig.~\ref{fig:NuEavg1}, but $10\,{\rm ms}$ after merger.}
\label{fig:NuEavg10}
\end{figure}

We can glean more information about the average neutrino energies by looking at the angular dependence of the estimate $\langle \nu \rangle$, shown in 
Figs.~\ref{fig:NuEavg1}-\ref{fig:NuEavg10} at $1\,{\rm ms}$, $5\,{\rm ms}$, and $10\,{\rm ms}$ after merger. 
Around merger, the general trend is for polar neutrinos to be of 
higher energy than equatorial neutrinos. There are, however, significant variations on
top of that general trend, as most of the emission comes from localized hot spots. The results are much clearer after disk formation. $5\,{\rm ms}$ after
merger, the
equatorial neutrinos are coming from the shocked tidal arms in the disk. As the disk material is mostly moving in the azimuthal direction, this results
in a large peak in the average neutrino energy in the equatorial direction $\Theta = 0$. This feature is less visible at the end of the simulation ($10\,{\rm ms}$
after merger), 
when the optical depth of the disk decreases and neutrinos emitted by fluid elements with a wider range of velocities contribute to the equatorial emission. Additionally,
as the disk expands, the velocity of the emitting fluid in the hot tidal arms decreases, which also contributes to a decrease in the effect of relativistic beaming in the
equatorial plane.
At all times, the core of the remnant is hotter than the disk, and dominates the
polar emission. Accordingly, the average neutrino energy increases in the $30^\circ-40^\circ$ cone around the poles in which most of the neutrinos are emitted.

The fact that this last effect is stronger for $\nu_e$ than for $\bar\nu_e$ has some consequences for the absorption of neutrinos by the polar
disk winds. Indeed, the leakage scheme overestimates the energy difference between the polar $\nu_e$ and $\bar\nu_e$. With the local estimate
of the average neutrino energy, the absorption of $\nu_e$ will increase, which is one of the factors contributing to less neutron rich outflows (see Sec.~\ref{sec:outflows}). 

Finally, we note that our method to evolve the neutrino number density, and in particular the computation of the neutrino number flux, involves the use of an
ad-hoc parameter $\beta$ (see Appendix), necessary to take into account the fact that low-energy neutrinos diffuse more easily through the fluid than
high-energy neutrinos. To test the impact of that free parameter at a reasonable computational cost, we perform short evolutions on a static
background (fixed metric and fluid density), corresponding to the final state of our simulation $10\,{\rm ms}$ after merger. As in the post-bounce core-collapse test problem presented in the Appendix,
we find that varying $\beta$ between the unphysical extremes $\beta\rightarrow 0$ and $\beta \rightarrow \infty$ causes changes of $10\%-20\%$ in the neutrino
luminosity and of $\lesssim 1\,{\rm MeV}$ in the average neutrino energies. Negligible differences are observed over the more physically realistic range $\beta = 4-8$.
The only exception is the predicted average energy of the heavy lepton neutrinos, which is heavily overestimated when $\beta \rightarrow \infty$ in both the test 
problem and the post-merger evolution ($\beta \rightarrow \infty$ is the limiting case in which the spectrum of the neutrino flux and energy density are assumed
to be the same). This is most likely a consequence of the large difference between scattering and absorption opacities for $\nu_x$. 
This is a particularly unfavorable regime for a gray transport scheme, and
exactly the regime for which we found the introduction of the free parameter $\beta$ to be necessary.
 
Overall, it thus appears that in the binary neutron star merger considered here, our improved estimate of the neutrino average energy has a noticeable
impact on: (1) the neutrino luminosity, particularly by decreasing the luminosity of $\bar\nu_e$ and $\nu_x$ by $\sim 30-40\%$; 
(2) the estimated average neutrino energy
at early times for all neutrino species, and at all times for $\nu_x$, generally increasing those energy estimates; (3) the 
spatial distribution of the average neutrino energy, which is now approximately captured (within the limits of the M1 closure) 
instead of being averaged over all optically thin regions; and (4) the conservation of the total lepton number in the simulation, which is exact with our new scheme. 
We can now study how these changes affect an observationally important aspect of the merger: the composition of the material ejected from the system 
during and after merger.

\section{Outflow properties}
\label{sec:outflows}

\begin{figure}
\flushleft
\includegraphics[width=1.0\columnwidth]{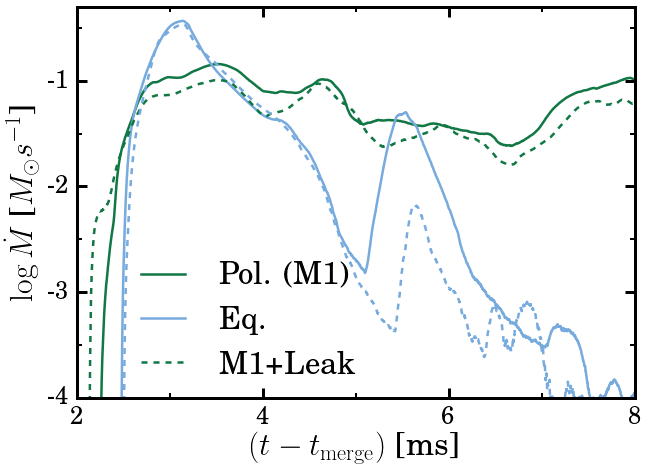}
\caption{Outflow rate of unbound material across the outer boundary of the computational domain for simulations using a local estimate of the average neutrino
energy (this work, solid lines) and a global estimate from a leakage scheme (from Paper I, dashed lines). The simulations have very similar 
outflow rates in both the polar (green curve) and equatorial (blue curve) directions. The peak in the equatorial ejection of material at $5.5\,{\rm ms}$
is mostly an effect of the cubical grid used in our simulations: the outflow rate is larger when the tidal tail reaches the center of a face of the computational
domain.}
\label{fig:MassOutflows}
\end{figure}

Neutrino-matter interactions set the composition
of the fluid both in the post-merger remnant and in the material ejected by the merger, which can strongly impact the observable signatures of neutron star mergers. The fluid composition in our simulations 
is described by the electron fraction
\beq
Y_e = \frac{n_p}{n_p+n_n},
\eeq
with $n_p$ and $n_n$ being the proton and neutron number densities, respectively (the net electron number density $n_{e^-}-n_{e^+}=n_p$, due to charge neutrality in the fluid).
Neutrino absorption in low-density regions can also drive winds of unbound material above and below the remnant accretion disk. 
Both of these effects are important to assess in order to estimate the properties of the material ejected
by the merger, and in particular of the transients observable in the optical and/or infrared bands as a consequence of rapid neutron capture (r-process) 
nucleosynthesis in the  ejecta~\cite{Kasen:2013xka,2013ApJ...775...18B}. For the typical entropy and velocity observed in neutron star mergers, in particular, nucleosynthesis in the ejecta can lead to two distinct outcomes. For neutron rich material, the r-process leads to the formation of heavy, neutron rich nuclei 
whose radioactive decay results in the production of stable elements with mass number $A \gtrsim 120$. 
In that case, nucleosynthesis yields are fairly independent of the initial conditions, and robustly match observed solar system abundances for  
$A \gtrsim 120$ (strong r-process) -- but not for lower mass elements also generally associated with r-process nucleosynthesis~\cite{korobkin:12}.
On the other hand, if the ejecta is less neutron rich, rapid neutron capture ends before the formation of heavy elements, and r-process nucleosynthesis results
instead in the formation of lower mass elements (weak r-process). 
Accordingly, neutrino-matter interactions driving up the electron fraction of the ejected material can play
a significant role in the relative production of low and high mass r-process elements~\cite{Wanajo2014,Sekiguchi:2015,Goriely:2015}. For the material ejected
in the mergers studied here, the threshold to avoid the strong r-process is $Y_e\approx 0.23$~\cite{Lippuner2015}, with potentially significant uncertainties due
to both unknown nuclear physics and the exact velocity and entropy of the ejecta.

A consequence of those different nucleosynthesis results is a drastic change in the optical opacity of the ejecta once r-process nucleosynthesis ends. 
Some heavy r-process nuclei have a particularly high-opacity, which is
expected to cause radiation from a more neutron-rich ejecta to peak in the infrared on a timescale
of a week~\cite{Kasen:2013xka,2013ApJ...775...18B}. If these heavy nuclei are not produced, however, the electromagnetic transient following
the merger should peak in the optical on a timescale of about a day.

In our simulations, we find that using an improved estimate of the average neutrino energy has an important impact on the predicted result of r-process nucleosynthesis
in the ejecta. There are two main components to the ejecta observed in our simulation: a cold, neutron rich equatorial ejecta coming from the tidal disruption of the
neutron stars, and a hot, polar ejecta coming from shocks at the time of merger, and neutrino-driven winds after merger. Fig.~\ref{fig:MassOutflows} shows the mass
outflow for both types of ejecta in simulations using a global estimate of the average neutrino energy, and with our improved local scheme. We see that the amount
of mass ejected by the merger is fairly similar in the two sets of simulations: at late times, mass loss in the polar region
is increased by $\sim 20\%$ in the simulation evolving the neutrino number density. This is to be contrasted with simulations neglecting neutrino absorption,
which do not show sustained polar outflows after merger (see Paper I)~\footnote{We note that the impact of the choice of an M1 closure on the properties
of these neutrino-driven outflows is uncertain. The M1 scheme is known to produce radiation shocks in the polar regions which may affect the results, but the exact
impact of those shocks cannot be tested without using a completely different transport scheme.}, 
unless one also takes into account magnetically-driven outflows~\cite{Kiuchi2014}.

\begin{figure}
\flushleft
\includegraphics[width=1.0\columnwidth]{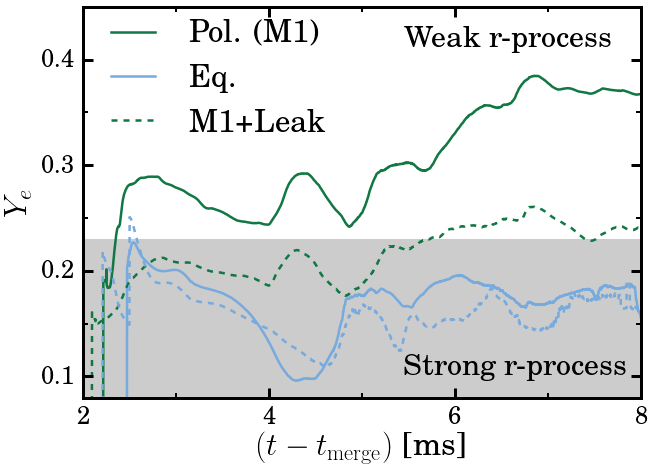}
\caption{Average electron fraction of the material leaving the computational domain for simulations using a local estimate of the average neutrino
energy (this work, solid lines) and a global estimate from a leakage scheme (from Paper I, dashed lines). The polar ejecta (green curve) is significantly 
less neutron rich when using the local estimate of the neutrino energy. The shaded gray region approximately covers the range of $Y_e$ over which we 
expect strong r-process nucleosynthesis in the ejected material. The equatorial ejecta (blue curve) is neutron-rich in both simulations.}
\label{fig:OutflowYe}
\end{figure}

\begin{figure*}
\center
\includegraphics[width=2.0\columnwidth]{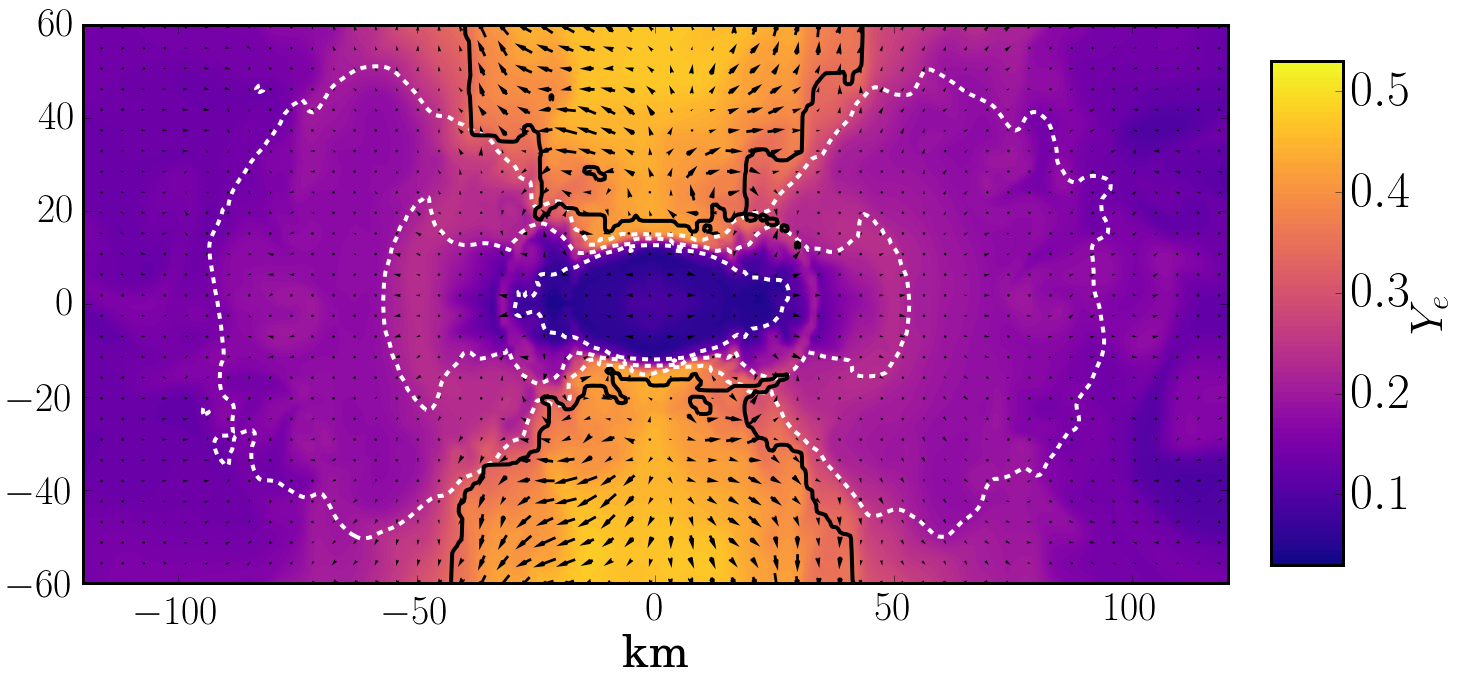}
\caption{Vertical slice through the numerical simulation $10\,{\rm ms}$ after merger. The color gradient shows the electron fraction of the fluid.
Dashed white lines show isodensity contours $\rho_0 = 10^{10,11,12}\,{\rm g\,cm^{-3}}$. Arrows show the transport velocity in the fluid. 
The solid black line shows the boundary of the region in which the fluid
is marked as unbound. All unbound material (i.e. fluid elements in the polar regions) has a high electron fraction $Y_e>0.25$.}
\label{fig:PolEjLate}
\end{figure*}

The main effect of the corrected average neutrino energies on the outflows can be observed on Fig.~\ref{fig:OutflowYe}. 
The electron fraction of the polar ejecta, which in previous simulations was hovering right around the dividing line between weak and strong 
r-process nucleosynthesis, is now clearly high enough to prevent strong r-process nucleosynthesis. The increase in the electron fraction
of the polar ejecta can easily be understood from the changes in the neutrino properties described in the previous section. In particular,
the higher energy of the electron neutrinos in the polar regions, combined with the lower number of electron antineutrinos, makes it much
easier to convert neutrons to protons through preferential absorption of electron neutrinos. We note that the change in electron fraction appears
to be the only important difference between the outflows in both simulations. We have already seen that the ejected mass is only modified by $\sim 20\%$.
The specific entropy of the polar wind changes even less, with an increase of $\sim 3\%$ in the entropy of the outflows in the simulation evolving the
neutrino number density (in the polar outflows, $s\sim 30k_B$, with $s$ the specific entropy per nucleon). 
This small difference in the entropy of the outflows indicates that the increase in $Y_e$ is due mostly to a change in the relative
number of $\nu_e$ and $\bar \nu_e$ absorptions in the outflows (or, equivalently, a change in the value of $Y_e$ at which the outflows
are in equilibrium with the neutrino radiation), rather than to additional absorptions of $\nu_e$ alone.

We note that the electron fraction of the polar outflows is largely set by neutrino emission and absorption very close
to the compact neutron star core, where the temperature of the fluid and the neutrino fluxes are the highest. In that region, the value
of $Y_e$ at which the fluid is in equilibrium with the neutrinos is $Y_e^{\rm eq}\sim 0.4-0.5$. Farther from the core, electron antineutrinos emitted
from the tidal arm contribute more significantly to the equilibrium composition, driving it down to $Y_e^{\rm eq}\sim 0.25-0.35$ (with the lower
values being observed at earlier times). This
explains the gradient of $Y_e$ in the low-density regions close to the compact remnant. In the equatorial regions, on the other
hand, there is a large excess of electron antineutrinos. There, the equilibrium composition is $Y_e^{\rm eq}\sim 0.1-0.2$, with the lower
values once more corresponding to earlier times. This indicates that, as opposed to what is observed in the polar regions, in the equatorial regions
neutrino absorption drives the fluid composition to values at which strong r-process nucleosynthesis is still expected.

The electron fraction of the ejected material is also large everywhere in the polar regions, not just on average. Fig.~\ref{fig:PolEjLate} shows the electron
fraction in a vertical slice of the computational domain, $10\,{\rm ms}$ after merger. All of the polar ejecta is at electron fractions $Y_e\gtrsim 0.25$,
which should be sufficient to avoid strong r-process nucleosynthesis.

The fact that neutrino absorption in the polar regions can increase the electron fraction of the ejecta has generally been observed in all general relativistic
simulations of post-merger remnants using an approximate neutrino transport scheme~\cite{Sekiguchi:2015,FoucartM1:2015,FoucartM1:2016,Sekiguchi:2016}.
Our results show that, in the gray approximation, the way in which we estimate the average energy of the neutrinos can have important consequences for
the magnitude of that effect. For the configuration studied in this work, evolving the neutrino number density to obtain a local estimate of the 
neutrino average energy makes it clear that the polar ejecta is initially prevented from undergoing strong r-process nucleosynthesis. This is a prerequisite
if we want to observe an early, optical counterpart to the merger. Indeed, this optical counterpart could be obscured by high-opacity lanthanides at the lower electron 
fractions predicted by our simulation using a single average energy in optically thin regions for each species of neutrinos. 
We note that on the other hand, in the equatorial regions,
the ejection of neutron-rich material is a robust feature of binary neutron star mergers. Optical emission could thus be visible if the merger is observed 
face-on, while only an infrared, longer lived emission can be visible when observing the merger edge-on.

This difference between polar and equatorial ejecta, and the lack of strong r-process in the former, is particularly important if high-$Y_e$ magnetically-driven winds
can be powered over a much longer timescale in the post-merger remnant~\cite{Kiuchi2014}. The long-term evolution of post-merger accretion disks has not provided
us with definitive answers as to whether a strong r-process robustly occurs in post-merger disk winds. 
Predicted electron fractions remain sensitive to the initial conditions
of the post-merger evolution and the included microphysics, and are generally close to the boundary between strong and weak r-process 
nucleosynthesis~\cite{Just2014,Perego2014,Metzger2014,Fernandez:2014,Fernandez:2014b}.

\section{Conclusions}
\label{sec:conclusions}

We have presented a detailed study of a NSNS merger with a general relativistic hydrodynamics code and two variations of an approximate, gray neutrino 
transport scheme. We considered in particular the impact of the method used to approximate the neutrino energy spectrum on the post-merger evolution of the system. 
In previous simulations (Paper I), we estimated the average neutrino energy in all optically thin regions using a single neutrino temperature for each neutrino species, 
taken from the  prediction of a simple leakage scheme~\cite{FoucartM1:2015,FoucartM1:2016}. In this work, we instead evolved the neutrino number and 
energy density, and used those evolved variables to estimate a spatially-varying average neutrino energy.

The new scheme has the advantages of exactly conserving the total lepton number, taking into account spatial variations in the neutrino energy, and being sensitive to the impact of relativistic beaming on the average neutrino energies. 
It generally predicts higher neutrino energies, 
particularly immediately after merger and in the polar regions, and neutrino luminosities differing by $\lesssim 40\%$. 

These differences do not appear to affect the dynamics of the post-merger remnant, or to have a significant impact on its temperature. However, they do have
important consequences for the evolution of the composition of the fluid. Material unbound in the polar regions as a neutrino-driven wind absorbs
fewer electron antineutrinos and more electron neutrinos when using local estimates of the average neutrino energy. This robustly drives the electron fraction of the
polar ejecta to values $Y_e \gtrsim 0.25$, with an increase of $\Delta Y_e \sim 0.05-0.1$ with respect to results using a global estimate of the average neutrino energy.
The low-density, bound regions of the remnant also see an increase in their electron fraction.

Such a change in the average electron fraction of the polar ejecta could have important consequences for the observable electromagnetic counterpart of the merger
due to r-process nucleosynthesis in the ejecta. In the absence of neutron-rich ejecta in the polar regions, the opacity of the ejecta along the line of sight of an
observer viewing the merger face-on could be significantly reduced. This makes it possible to observe electromagnetic transients peaking in the optical when
the merger is observed face-on, particularly if high-$Y_e$ disk winds continue to be ejected by the post-merger remnant over timescales significantly longer than
the duration of our simulation~\cite{Just2014,Perego2014,Metzger2014,Fernandez:2014,Fernandez:2014b}.

Although we believe that the new methods presented here provide a more accurate representation of the merger than our previous results (Paper I),
the strong dependence of the polar electron fraction on the method used to estimate the average neutrino energy may offer us a first view of the limits of current gray
neutrino transport schemes. After all, even our improved estimate of the neutrino spectrum remains fairly rudimentary. The difference between the composition of the
ejecta in the two neutrino transport schemes is only slightly smaller than the difference observed in Paper I between a leakage scheme ignoring neutrino absorption and
the neutrino transport schemes. It may thus be useful to obtain better predictions for the neutrino energy spectrum in the polar regions.
We should also note that the moment formalism used here to approximate the neutrino distribution function is notoriously problematic in regions in which
radiation beams emitted from different directions cross paths. This is obviously the case in large parts of the polar regions, where most of the neutrino-matter
interactions that drive up the electron fraction of the wind take place. 
Even if the neutrino-matter interactions are reasonably well approximated by the current scheme in a volume-averaged sense,
the exact impact of the moment formalism on the neutrino emission and properties of the outflows remain an important question for future studies of binary
neutron star mergers.

\acknowledgments
The authors thank Matthew Duez, Dan Hemberger, and the members of the SxS collaboration for their input and support 
during this project, Dan Kasen and Rodrigo Fernandez for regular discussions on binary mergers and outflows,
and Brett Deaton for his comments on an earlier version of this manuscript.
Support for this work was provided
by NASA through Einstein Postdoctoral Fellowship
grants numbered PF4-150122 (F.F.) and PF3-140114 (L.R.) awarded 
by the Chandra X-ray Center, which is operated by the Smithsonian 
Astrophysical Observatory for NASA under contract NAS8-03060; 
and through Hubble Fellowship grant number 51344.001 awarded 
by the Space Telescope Science Institute, which is operated by the Association
 of Universities for Research in Astronomy, Inc., for NASA, under contract NAS 5-26555.
The authors at CITA gratefully acknowledge support from the NSERC
Canada. 
L.K. acknowledges support from NSF grants PHY-1306125 and AST-1333129 at
Cornell, while the authors at Caltech acknowledge support from NSF
Grants  PHY-1404569, AST-1333520, NSF-1440083, and
NSF CAREER Award PHY-1151197.
Authors at both Cornell and Caltech also thank the Sherman Fairchild Foundation
for their support. 
Computations were performed on the
supercomputer Briar\'ee from the Universit\'e de Montr\'eal, and Guillimin
from McGill University, both managed
by Calcul Qu\'ebec and Compute Canada. The operation
of these supercomputers is funded by the Canada Foundation
for Innovation (CFI), NanoQu\'ebec, RMGA and the Fonds de
recherche du Qu\'ebec - Nature et Technologie (FRQ-NT). Computations were
also performed on the Zwicky cluster at Caltech, supported by the Sherman
Fairchild Foundation and by NSF award PHY-0960291.
This work also used the Extreme Science and Engineering
Discovery Environment (XSEDE) through allocation No. TGPHY990007N,
supported by NSF Grant No. ACI-1053575.

\bibliography{References/References}

\appendix
\section{Improved gray two-moment formalism for neutrino transport}
\label{sec:form}

We now describe in more detail our improvements to the two-moment
scheme for neutrino transport used by the SpEC code,
and in particular the evolution within this scheme of the neutrino number density.
The number density provides us with additional information about the average energy of neutrinos
at any given point. 

We first define the moments of the neutrino
distribution function in Sec.~\ref{sec:mom}, then provide an overview of
the evolution equations for those moments in Sec.~\ref{sec:M1eq} 
and of their implementation in SpEC in Sec.~\ref{sec:M1impl}.

The M1 transport scheme presented in those sections has
a number of undefined variables, for which some reasonable approximations
have to be implemented in order to close the system of equations. These include
the average energy of the neutrino flux, discussed in Sec.~\ref{app:flux},
and the energy-integrated source terms and energy-averaged opacities,
discussed in Sec.~\ref{sec:M1source}. While we attempt to make reasonable
choices for these variables, it should be acknowledged that there is no
truly correct way to define them within the moment formalism: in practice,
knowledge of higher moments of the distribution function of the neutrinos
would be required to properly compute them. The choices made here should
thus be seen as part of a partially motivated but nonetheless very approximate
closure. An energy-dependent scheme would be necessary to get rid of 
these approximations.

\subsection{Gray Moments}
\label{sec:mom}

For each neutrino species $\nu_i$, we can describe
the neutrinos by their distribution function
$f_{(\nu)}(x^\alpha,p^\alpha)$, where $x^\alpha=(t,x^i)$
gives the time and the position of the neutrinos,
and $p^\alpha$ is the 4-momentum of the neutrinos.
The distribution function $f_{(\nu)}$ evolves according to the 
Boltzmann equation
\beq
p^\alpha \left[\frac{\partial f_{(\nu)}}{\partial x^\alpha} - \Gamma^\beta_{\alpha \gamma}
p^\gamma \frac{\partial f_{(\nu)}}{\partial p^\beta} \right] = \left[\frac{d f_{(\nu)}}{d\tau}\right]_{\rm coll}\,\,,
\eeq
where the $\Gamma^\alpha_{\beta \gamma}$ are the Christoffel symbols and the right-hand side includes all collisional
processes (emissions, absorptions, scatterings). In general, this is a 7-dimensional problem
which is extremely expensive to solve numerically. Approximations to the Boltzmann equation
have thus been developed for numerical applications. In this work, we consider the moment
formalism developed by Thorne~\cite{1981MNRAS.194..439T}, in which only the lowest moments
of the distribution function in momentum space are evolved. 

We use this formalism in the ``gray'' approximation, that is we only consider energy-integrated moments. 
Although the moment formalism can in theory be used with a discretization in neutrino energies, this makes the
simulations significantly more expensive and involves additional technical difficulties in the treatment
of the gravitational and velocity redshifts, particularly for applications such as compact binary mergers in which
we have both relativistic speeds and large gravitational redshifts. 
We consider three independent neutrino
species: the electron neutrinos $\nu_e$, the electron antineutrinos $\bar \nu_e$, and the heavy-lepton
neutrinos $\nu_x$. The latter is the combination of 4 species
($\nu_\mu, \bar\nu_\mu,\nu_\tau,\bar\nu_\tau$).  This merging is
justified because the temperatures and neutrino energies reached in
our merger calculations are low enough to suppress the formation of
the corresponding heavy leptons whose presence would require including
the charged current neutrino interactions that differentiate between
these individual species.

In the gray approximation, and considering only the first two moments of the distribution function,
we evolve for each species projections of the
stress-energy tensor of the neutrino radiation $T^{\mu \nu}_{\rm rad}$. 
One possible decomposition of $T^{\mu \nu}_{\rm rad}$ is~\cite{shibata:11}
\beq
T^{\mu \nu}_{\rm rad} = J u^\mu u^\nu + H^\mu u^\nu + H^\nu u^\mu + S^{\mu \nu}\,\,,
\eeq
with $H^\mu u_\mu = S^{\mu \nu}u_\mu = 0$ and $u^\mu$ the 4-velocity of
the fluid. The energy $J$, flux $H^\mu$
and stress tensor $S^{\mu \nu}$ of the neutrino radiation as observed by an
observer comoving with the fluid are related to the neutrino distribution function by
\beqn
J &=& \int_0^\infty d\nu\,\nu^3 \int d\Omega\,f_{(\nu)}(x^\alpha, \nu,\Omega)\,\,,\\
H^\mu &=& \int_0^\infty d\nu\,\nu^3 \int d\Omega\,f_{(\nu)}(x^\alpha,\nu,\Omega) l^\mu\,\,,\\
S^{\mu\nu} &=& \int_0^\infty d\nu\,\nu^3 \int d\Omega\,f_{(\nu)}(x^\alpha,\nu,\Omega) l^\mu l^\nu\,\,,
\eeqn
where $\nu$ is the neutrino energy in the fluid frame, $\int d\Omega$ denotes integrals
over solid angle on a unit sphere in momentum space, and
\beq
p^\alpha = \nu (u^\alpha + l^\alpha)\,\,,
\eeq
with $l^\alpha u_\alpha = 0$ and $l^\alpha l_\alpha =1$.
We also consider the decomposition of $T^{\mu \nu}_{\rm rad}$ in
terms of the energy, flux and stress tensor observed by an
inertial observer,
\beq
T^{\mu \nu}_{\rm rad} = E n^\mu n^\nu + F^\mu n^\nu + F^\nu n^\mu + P^{\mu \nu}\,\,,
\eeq
with $F^\mu n_\mu=P^{\mu \nu} n_\mu = F^t = P^{t\nu}=0$, 
and $n^\alpha$ the unit normal to a $t={\rm constant}$ slice. Additionally, we consider
for each species the number current density of neutrinos,
\beq
N^\mu = N n^\nu + \mathcal{F}^\mu
\eeq
with $N$ the number density of neutrinos, and $\mathcal{F}^\mu$ the number flux density.
In a previous implementation of the moment formalism as a gray scheme (Paper II~\cite{FoucartM1:2015}), 
we only evolved
$E$ and $F_i$. Whenever information about the neutrino spectrum was required, we then
either assumed a black-body distribution function at the temperature of the fluid (in
optically thick regions), or used a global estimate of the average neutrino energy from
an approximate leakage scheme (in optically thin regions). To improve on this
method, and obtain a local estimate of the average neutrino energy everywhere,
we now consider an algorithm in which for each neutrino species we evolve the
variables $(N,E,F_i)$. This algorithm also has the advantage of decoupling the transport
scheme from the leakage scheme, and of consistently keeping track of the total lepton number. 

From the number current density, we define the average neutrino energy in the fluid frame
$\langle \nu \rangle$ through the equation
\beq
N^\mu = \frac{J u^\mu + H^\mu}{\langle \nu \rangle}.
\eeq
If we decompose the 4-velocity as
 \beq
u^\mu = W (n^\mu + v^\mu)\,\,,
\eeq
with $v^\mu n_\mu=0$ and $W$ the Lorentz factor, we can get the alternate
expression
\beq
\langle \nu \rangle = W \frac{E - F_i v^i}{N},
\eeq
where we have used the identity
\beq
T^{\mu \nu} u_\mu n_\nu = JW - H^\mu n_\mu = EW - F^\mu u_\mu.
\eeq
An important assumption in our algorithm is the choice of the form
of the neutrino distribution function $f_{(\nu)}$.
We generally assume a blackbody spectrum. We then have
\beq
f_{(\nu)}^{BB} = \frac{1}{1+\exp[(\nu-\mu_\nu)/T_\nu]},
\eeq
or, in terms of the energy density,
\beq
E_{(\nu)} \propto \frac{\nu^3}{1+\exp[(\nu-\mu_\nu)/T_\nu]},
\eeq
with $\mu_\nu$ the chemical potential of neutrinos in equilibrium with
the fluid. Defining the Fermi integrals
\beq
F_k(\eta_\nu) = \int_0^{\infty} \frac{x^k}{1+\exp{(x-\eta_\nu)}}\,dx,
\eeq
we get the relationship between the neutrino temperature and average energy
for a black body spectrum
\beq
\langle \nu \rangle = \frac{F_3(\eta_\nu)}{F_2{(\eta_\nu})} T_\nu,
\eeq
with $\eta_\nu = \mu_\nu/T$.
By evolving $N$, we can now hope to get reasonable estimates of $T_\nu$
everywhere, in a sense that will be discussed in more detail below.

\subsection{Evolution Equations}
\label{sec:M1eq}

The evolution equations are very similar to those used in our previous algorithm.
We use the 3+1 decomposition of the metric,
\beqn
ds^2 &=& g_{\alpha \beta}dx^\alpha dx^\beta\\
&=& -\alpha^2 dt^2 + \gamma_{ij} (dx^i + \beta^i)(dx^j+\beta^j)\,\,,
\eeqn
where $\alpha$ is the lapse, $\beta^i$ the shift, and $\gamma_{ij}$
the 3-metric on a slice of constant coordinate $t$. The extension
of $\gamma_{ij}$ to the full 4-dimensional space is the projection
operator 
\beq
\gamma_{\alpha \beta} = g_{\alpha \beta} + n_\alpha n_\beta\,\,.
\eeq
We similarly define a projection operator onto the reference frame 
of an observer comoving with the fluid, 
\beq
h_{\alpha \beta}=g_{\alpha \beta}+u_\alpha u_\beta\,\,.
\eeq
We can then
write equations relating the fluid-frame variables to the inertial frame variables~\cite{Cardall2013}:
\beqn
E &=& W^2 J + 2W v_\mu H^\mu + v_\mu v_\nu S^{\mu \nu}\,\,,\\
F_\mu &=& W^2 v_\mu J + W (g_{\mu\nu}-n_\mu v_\nu)H^\nu \nonumber\\
&& +Wv_\mu v_\nu H^\nu + (g_{\mu\nu}-n_\mu v_\nu)v_\rho S^{\nu
\rho} \label{flux_equation} \,\,,\\
P_{\mu \nu} &=& W^2 v_\mu v_\nu J + W (g_{\mu\rho}-n_\mu v_\rho)v_\nu H^\rho \nonumber\\
&&+W(g_{\rho\nu}-n_\rho v_\nu)v_\mu H^\rho \nonumber\\
&&+(g_{\mu\rho}-n_\mu v_\rho)(g_{\nu\kappa}-n_\nu v_\kappa)S^{\rho \kappa}\label{eq:Pij}\,\,.
\eeqn

Evolution equations
for $\tilde E = \sqrt{\gamma}E$, $\tilde F = \sqrt{\gamma}F^i$, and $\tilde{N}=\sqrt{\gamma}N$
can then be written in conservative form:
\beqn
\partial_t \tilde E &+& \partial_j(\alpha \tilde F^j -\beta^j \tilde E)\label{eq:Enu}\\
&=&\alpha (\tilde P^{ij}K_{ij} -\tilde F^j \partial_j \ln{\alpha} - \tilde S^\alpha n_\alpha)\nonumber\,\,,\\
\partial_t \tilde F_i &+& \partial_j(\alpha \tilde P^j_{i} -\beta^j \tilde F_i) \label{eq:Fnu}\\
&=&(-\tilde E\partial_i\alpha+\tilde F_k\partial_i\beta^k+\frac{\alpha}{2} \tilde P^{jk}\partial_i \gamma_{jk}+\alpha \tilde S^\alpha \gamma_{i\alpha})\nonumber\,\,,\\
\partial_t \tilde N &+& \partial_j(\alpha \sqrt{\gamma} \mathcal{F}^j -\beta^j \tilde N) = \alpha \sqrt{\gamma}C_{(0)}\label{eq:Nnu} \,\,
\eeqn
where $\gamma$ is the determinant of $\gamma_{ij}$, and $\tilde P_{ij}=\sqrt{\gamma} P_{ij}$.

To close this system of equations, we need three additional ingredients: a prescription for the 
computation of $P^{ij}(E,F_i)$ (`closure relation', which we choose following Minerbo 1978~\cite{Minerbo1978}), 
a prescription for the computation of the number flux $\mathcal{F}^i$ (specific to the evolution of the number density
$N$ in this paper and described in more detail in Sec.~\ref{sec:M1impl}),
and the collisional source terms ($\tilde S^\alpha$, $C_{(0)}$). In the M1 formalism, the neutrino pressure tensor $P^{ij}$ 
is recovered as an interpolation between its known limits for an optically thick medium and an optically thin medium 
with a unique direction of propagation for the neutrinos. Details on its computation are available in Paper II.
For the source terms, we will consider that the fluid has an energy-integrated emissivity $\bar\eta$ due
to the charged-current reactions
\beqn
p + e^- &\rightarrow& n + \nu_e\,\,,\\
n + e^+ &\rightarrow& p + \bar \nu_e\,\,,
\eeqn
as well as electron-positron pair annihilation
\beq
e^+ + e^- \rightarrow \nu_i \bar\nu_i\,\,,
\eeq 
plasmon decay
\beq
\gamma \rightarrow \nu_i \bar\nu_i\,\,,
\eeq
and nucleon-nucleon Bremsstrahlung
\beq
N + N \rightarrow N + N + \nu_i + \bar \nu_i\,\,.
\eeq
The inverse reactions are responsible for an energy-averaged absorption opacity
$\bar\kappa_a$. We also consider an energy-averaged scattering opacity $\bar\kappa_s$ due to elastic scattering
of neutrinos on nucleons and heavy nuclei. The source terms $\tilde S^\alpha$ are then
\beq
\tilde S^\alpha = \sqrt{\gamma} \left(\bar\eta u^\alpha - \bar\kappa_a J u^\alpha - (\bar\kappa_a+\bar\kappa_s) H^\alpha\right)\,\,.
\eeq
We use the emissivities and opacities proposed by Ruffert et al.~\cite{Ruffert1996} for all of the above reactions, except for
nucleon-nucleon Bremsstrahlung for which the emissivity is computed following Burrows et al.~\cite{Burrows2006b}.
The collisional source term for the number density $\tilde N$ is given by
\beq
C_{(0)} = \bar{\eta}_N - \bar{\kappa}_N \frac{J}{\langle \nu \rangle} = \bar{\eta}_N - 
\frac{\bar{\kappa}_N J \tilde{N}}{W(\tilde{E}-\tilde{F}_iv^i)},
\eeq
with $\bar{\eta}_N$ the energy-integrated number emission and $\bar{\kappa}_N$ the energy-averaged number absorption.
Properly choosing the relationship between the source terms $(\bar{\eta}_N,\bar{\kappa}_N,\bar{\eta},\bar{\kappa}_A,\bar{\kappa}_S)$ are important steps in obtaining reasonable estimates of the average neutrino energy, discussed in more detail in Sec.~\ref{sec:M1source}.

\subsection{Numerical scheme}
\label{sec:M1impl}

We add the evolution of neutrinos with the moment scheme to the SpEC code~\cite{SpECwebsite}, which
already includes a general relativistic hydrodynamics module~\cite{Duez:2008rb}. The latest methods
used for evolving in SpEC the coupled system formed by Einstein's equation and the general relativistic equations of hydrodynamics 
are described in~\cite{Foucart:2013a}, Appendix A. The basic steps used to evolve the moments of the neutrino distribution functions
were outlined in Paper II. Here, we only focus on aspects specific to the addition of the number density $\tilde N$.

An advantage of evolving $\tilde N$ is that the change in the composition of the fluid can now be computed very simply.
We have 
\beq
\partial_t (\rho_* Y_e) = ...\, - {\rm sign}(\nu_i) \alpha \sqrt{\gamma}C_{(0)}\label{eq:sourceRhoYe}\,\,,
\eeq
where $\rho_*$ is the conserved variable
\beq
\rho_* = \rho_0 W \sqrt{\gamma} \,\,,
\eeq
$\rho_0$ is the baryon density of the fluid, $Y_e$ its electron fraction, and $\rm{sign}(\nu_i)$ is $1$ for $\nu_e$, $-1$ for $\bar\nu_e$,
and $0$ for heavy-lepton neutrinos. Evolving $\tilde N$ frees us from having to guess at the average neutrino energy when computing
the coupling to the fluid. It also guarantees that the source term for the evolution of the electron fraction of the fluid is fully consistent with
the evolution of the neutrino number density, thus conserving the total lepton number of the system. When $\tilde N$ is not evolved, as in
Paper I and Paper II, the total lepton number is not consistently evolved.
The energy and momentum source terms are not modified when evolving $\tilde N$.

As we evolve $\tilde N$, we now also have to compute the flux $F_N = \alpha \sqrt{\gamma} \mathcal F^i - \tilde N \beta^i$ at cell faces, 
and then take its divergence. We do so by reconstructing a left state and right state of the variables $(E,F/E,N/E)$ at 
cell faces from their value at cell centers, using shock-capturing reconstruction methods (in this work, MC). 
When computing $F_N$, we use the equality
\beq
\mathcal F^i = \frac{J W v^i}{\langle \nu \rangle} + \frac{\gamma^i_\mu H^\mu}{\langle \nu^F \rangle}
\eeq
with the average neutrino energy $\langle \nu \rangle$ computed from the reconstructed fields, and a correction to the
average energy of the neutrino flux $\langle \nu^F \rangle$ can be included. We describe in Sec.~\ref{app:flux} our choice of 
$\langle \nu^F \rangle$, made to take the effects of a finite optical depth on the spectrum into account.
We then combine these left and right states into a single face value $\bar F_N$ using the HLL Riemann solver,
\beq
\bar F_N = \frac{c_+ \bar F_{N,L} + c_- \bar F_{N,R} - c_+ c_- (\tilde N_R-\tilde N_L)}{c_+ + c_-}\,\,,
\eeq
where $c_+$ and $c_-$ are the absolute values of the largest right- and left-going characteristic speeds of the evolution
system (or zero if there is no left/right going characteristic speeds), as given in Paper II. The suffix $(R,L)$ denotes the left an
right state of the flux and number density.

As discussed in Paper II, in the optically thick limit these fluxes do not properly reproduce the diffusion rate of the neutrinos 
through the fluid. To recover the proper diffusion rate, we correct the energy density flux $\bar F_E$ \cite{Jin1996}
\beq
\bar F_{E,\rm corr} = a\bar F_E+(1-a) \bar F_{E,\rm asym}\,\,,
\eeq
with
\beqn
a &=& \tanh{\frac{1}{\bar \kappa \Delta x^d}}\,\,,\\
\bar \kappa_{i+1/2} &=& \sqrt{(\bar\kappa_a+\bar\kappa_s)_i (\bar\kappa_a+\bar\kappa_s)_{i+1}} \label{eq:ktface}\,\,,
\eeqn
and where half-integer indices refer to values of the opacities at cell faces while integer indices refer
to value of the opacities at cell centers. Here, $d$ is the direction in which we are reconstructing,
$\Delta x^d=\sqrt{g_{dd} (\Delta x^d_{\rm grid})^2}$ is the proper distance between two grid points along that direction,
and $\Delta x^d_{\rm grid}$ is the coordinate grid spacing along that direction.
The asymptotic flux in the fluid rest frame, which corresponds to
the flux in the diffusion limit, is \cite{1981MNRAS.194..439T} 
\beqn
H^{\rm asym}_\alpha = -\frac{1}{3 \bar\kappa} \partial_\alpha J_{\rm thick}\,\,,
\eeqn  
with $J_{\rm thick}$ computed assuming the optically thick closure relation $S^{\mu\nu}=(J/3) h^{\mu\nu}$. 
For consistency, we apply the same correction to the number density flux $\bar F_N$,
\beq
\bar F_{N,\rm corr} = a\bar F_N+(1-a) \bar F_{N,\rm asym}\,\,,
\eeq
with $F_{N,\rm asym}$ computed assuming $J=J_{\rm thick}$, $H^\mu = H^\mu_{\rm asym}$.
The numerical methods used to compute $H^\mu_{\rm asym}$ are described in detail in Paper II.
We only note that we use an upwind computation of $\partial_\alpha J_{\rm thick}$, and thus for consistency
use the upwind value of $\langle \nu^F \rangle$ (i.e. its value at the neighboring cell center) when computing
$(H^\mu_{\rm asym}/\langle \nu^F \rangle)$.

Finally, when evolving $\tilde N$, we treat the absorption term in $C_{(0)}$ implicitly, but all other terms explicitly.
We note that as $\tilde{N}$ does not appear in the evolution of $\tilde E$ or $\tilde F_i$, we can use operator splitting
to first evolve ($\tilde E$,$\tilde F_i$), and then evolve $\tilde N$ using the evolved values of  ($\tilde E$,$\tilde F_i$)
in $C_{(0)}$. 

\subsection{Energy of the neutrino flux}
\label{app:flux}

In Sec.~\ref{sec:M1impl}, we left the average energy of the neutrino flux, $\langle \nu^F \rangle$, undetermined.
The uncertainty in the determination of $\langle \nu^F \rangle$ is an important
limitation of the gray scheme used in this work. The choice of $\langle \nu^F \rangle$ is fairly unimportant in regions of high absorption opacity,
where the neutrinos remain in equilibrium with the fluid, or in regions of low optical depth, where $\langle \nu^F \rangle \approx \langle \nu \rangle$.
Regions of high scattering optical depth but low absorption optical depth are however problematic. As the opacities are steep functions of the neutrino energies, the spectrum of the neutrino flux can be significantly biased towards lower neutrino energies,
with $\langle \nu^F \rangle < \langle \nu \rangle$. Ignoring this effect can lead to significant overestimates of the neutrino energies in systems in which the scattering neutrinosphere is well outside of the absorption neutrinosphere, as well as underestimates of the diffusion rate of the neutrino number density.

Within the gray scheme, we cannot self-consistently take this effect into account. Instead, we rely on a simple parametrized model to include the first order effect of a large scattering region on $\langle \nu^F \rangle$. Given the ad-hoc nature of this model, any dependence of the numerical results on the parameters of the model is a sign that an energy-dependent treatment of the neutrinos may be necessary to obtain reliable results.

The starting point from our model is the fact that a blackbody spectrum of temperature $T_\nu$ 
going through a screen of high-opacity material with opacity proportional to $\nu^2$ (as it the case for the dominant neutrino-matter opacities) sees its average energy go from
$\langle \nu^{BB}\rangle = F_3(\eta_\nu)T_\nu/F_2({\eta_\nu})$ down 
to $\langle \nu^{sc}\rangle = F_1(\eta_\nu)T_\nu/F_0({\eta_\nu})$. We then make the choice
\beq
\frac{\langle \nu^{F}\rangle}{\langle \nu \rangle} = \frac{F_3F_0-s^F(F_3F_0-F_2F_1)}{F_3F_0-s^C(F_3F_0-F_2F_1)},
\label{eq:nuf}
\eeq
where, for simplicity, we have dropped the argument of the Fermi integrals. $0<s^C<1$ is a scalar representing the fraction of neutrinos which have gone through a significant optical depth since emission at a given point, and thus have a softer energy spectrum. $s^C<s^F<1$  is a scalar allowing
us to reduce the average energy of the neutrino flux in regions where $s^C\ll 1$, which effectively represents the fraction of neutrinos
which have gone through a significant optical depth in the neutrino flux. A simple choice for $s^C$ is
\beq
s^C = \frac{N s^C_0 + \alpha \mathcal F s^F dt}{N+ \alpha \mathcal F dt+\bar\eta_N \alpha dt } 
\label{eq:sc}
\eeq
with $s^C_0$ the value of $s^C$ at the beginning of the time step and $\mathcal F$ an estimate of the number flux of neutrinos at the given point.
The general idea behind this choice is to drive $s^C$ towards 0 in optically thick regions, under the assumption that neutrinos in those regions
are in equilibrium with the fluid and follow a black body distribution function, and to drive $s^C$ towards $s^F$ in optically thin regions, where the
neutrino spectrum is given by the spectrum of inflowing neutrinos. The ratio of inflowing neutrinos $\mathcal F$ to locally emitted neutrinos $\bar\eta_N$
appears to be a logical parameter to perform the transition between those two extremes (although this is clearly an approximation made
because of our lack of knowledge of the exact energy spectrum of the neutrinos).
For $s^F$, we make the choice
\beq
s^F = \frac{s^C + \tau}{1+\tau}
\label{eq:sf}
\eeq
with the optical depth $\tau$ approximated as 
\beq
\xi = \frac{1}{1+\beta \tau},
\eeq
$\xi = (H/J)$ the closure parameter, which we already compute to determine the neutrino pressure tensor $P^{ij}$,
and $\beta$ an arbitrary free parameter of the model. With this choice, $s^F\rightarrow 1$ in optically thick regions,
and $s^F \rightarrow s^C$ in optically thin regions, as desired. Keeping track of $s^C$ is necessary to avoid continually decreasing the average energy of neutrinos going through a large region of strong scattering: at most, the average energy of a packet
of neutrino emitted at temperature $T_\nu$ in an optically thick regions will drop from 
$\langle \nu^{BB}\rangle$ to $\langle \nu^{sc}\rangle$. The model thus reproduces some important effects of scattering regions on the average neutrino energy. 

The extreme choices $\beta\rightarrow 0$ and $\beta \rightarrow \infty$ correspond respectively to the simple choices $s^F = 1$ and $s^F=s^C$ everywhere, but neither of those choices can capture the effect of scattering regions on $\langle \nu \rangle$. Inspection of numerical solutions of NS-NS mergers, BH-NS mergers, and core-collapse supernovae using a leakage scheme in which $\tau$ is explicitly computed indicates that $\xi \sim 0.2$ on the neutrinosphere $\tau\sim 2/3$, leading us to the choice $\beta = 6$, but the choices $\beta \sim 4-8$ could be equally well justified.
We note that even the simple choice $\beta \rightarrow \infty$ ($s^F=s^C$ and $\langle \nu \rangle = \langle \nu^F \rangle$) can lead to significant
differences with the leakage-based scheme for the computation of the neutrino average energy used in Paper I and Paper II. The new scheme explicitly conserves
the total lepton number, provides a different estimate for the neutrino temperature than the leakage scheme, and accounts for velocity and gravitational redshifts
and relativistic beaming. Yet, the choice $\beta \rightarrow \infty$ leads to very inaccurate estimates of the average neutrino energies in the presence of a large
scattering region, as shown in Sec.~\ref{sec:test}.

To close the model, we now only need the approximate flux $\mathcal F$, for which we choose
\beq
\mathcal F = \xi N \left(\frac{F_3 F_0 -s^F_0 (F_3F_0-F_2F_1)}{F_2F_0} \right)^2.
\label{eq:Fcal}
\eeq
The first part of this equation, $ \xi N$, simply accounts for the ratio between the energy density and the energy flux in the
fluid frame. The last part is a purely ad-hoc correction for the fact that $\mathcal F/N > H/J$ in high opacity regions. 
The coefficient $s^F_0$ is the value of $s^F$ at the beginning of the time step (to avoid making~\ref{eq:Fcal}
an implicit equation).

The complete scheme to compute $\langle \nu^F\rangle$ is thus to first get the approximate optical depth
$\tau$ from $\xi$ and $\beta$, and the approximate flux $\mathcal F$ from~\ref{eq:Fcal}. Equations~\ref{eq:sc}
and~\ref{eq:sf} can then be combined into a simple linear equation for $s^C$. We then compute $s^F$ from~\ref{eq:sf}
and finally $\langle \nu^F\rangle$ from~\ref{eq:nuf}. At the first time step, we set $s^C=0$ everywhere, and $s^C$ very
rapidly evolve to its equilibrium value. We emphasize once more that the whole scheme is devise
to provide some reasonable estimate of $\langle \nu^F\rangle$ capturing the effect of a large scattering region.
While it improves on more primitive estimates for $\langle \nu^F\rangle$ (see Sec.~\ref{sec:test}), it is in no way
a replacement for a true energy-dependent scheme.

\subsection{Source Terms}
\label{sec:M1source}

The last missing component to allow us to evolve the moment of the neutrino distribution functions
are the energy-integrated emissivities $(\bar \eta,\bar \eta_N)$ and energy-averaged opacities 
$(\bar\kappa_A,\bar\kappa_S,\bar\kappa_N)$. Both play an important role in our updated scheme. In particular, the relation
between the energy and number emissivities/absorptions will determine our estimate of the average neutrino energy.
We first compute the energy-averaged absorption $\bar \kappa_A^{\rm eq}$ of charge-current processes, the energy-integrated emissivity $\bar \eta^{\rm eq}$ of thermal processes, and the energy-averaged scattering opacities $\bar \kappa_S^{\rm eq}$
for neutrinos in equilibrium with the fluid. We use the emissivities and opacities proposed by Ruffert et al.~\cite{Ruffert1996} for all reactions, except for nucleon-nucleon Bremsstrahlung for which the emissivity is computed following Burrows et al.~\cite{Burrows2006b}. We can then compute the equilibrium absorption opacities of charged current reactions and emissivities
of thermal processes using Kirchoff's law
\beq
\bar\eta^{\rm eq} = \bar\kappa^{\rm eq} \int B_{(\nu)}(T,\mu_\nu) d\nu,
\eeq
where $B_{(\nu)}$ is the blackbody spectrum at the fluid temperature $T$ for an equilibrium neutrino potential $\mu_\nu$.
Making use of the fact that the processes computed here have cross-sections scaling as $T_\nu^2$, we then make the choices
\beqn
\bar \eta &=& \bar \eta^{\rm eq}\,\,,\\
\bar \kappa_A &=& \bar \kappa_A^{\rm eq} \frac{T_\nu^2}{T^2}\,\,,\\
\bar \kappa_S &=& \bar \kappa_S^{\rm eq} \frac{T_\nu^2}{T^2}\,\,,
\eeqn
where $T_\nu^2$ is computed from the neutrino energy and number density, assuming a blackbody spectrum.
We can also choose the number absorption opacities so that the neutrinos are thermalized when the optical depth to absorption
is large (or, more precisely, as long as $\sqrt{\bar\kappa_A \bar\kappa_S}\gg1$). We simply need to set
\beq
\bar \kappa_N = \bar \kappa_A \frac{\bar\eta_N}{\bar\eta} \frac{F_3(\eta_\nu)T}{F_2(\eta_\nu)}\label{eq:kN},
\eeq
so that $T_\nu=T$ when $E=\bar\eta/\bar\kappa_a$, $N=\bar\eta_N/\bar\kappa_N$.
From Ruffert et al.~\cite{Ruffert1996} and Burrows et al.~\cite{Burrows2006b}, we also know the equilibrium number emission
$\bar \eta_N^{\rm eq}$, which we use to choose the last free source term $\bar\eta_N=\bar\eta_N^{\rm eq}$. 

\subsection{Test Problem: Spherically Symmetric Post-Bounce Supernova Profile}
\label{sec:test}

\begin{table}
\caption{
Neutrino luminosity (in units of $10^{51}{\rm erg/s}$) and energy-weighted neutrino energy $\langle \epsilon \rangle$ (in MeV) 
in one octant of the post-bounce supernova profile test, $8\,{\rm ms}$ after the beginning of the evolution. We show results for the
energy-dependent M1 scheme (with 12 energy groups) "Spectral M1", for our current gray M1 scheme with various choices
of the parameter $\beta$ (smaller values of $\beta$ imply a larger difference between the average neutrino energy in the flux
density and energy density in high opacity regions), and for the leakage scheme of~\cite{Deaton2013}.
 }
\label{tab:collapseprof}
\begin{ruledtabular}
\begin{tabular}{|c|c|c|c|c|c|c|} 
{\rm Scheme} & $L_{\nu_e}$ & $L_{\bar\nu_e}$ & $L_{\nu_x}$ & $\langle \epsilon_{\nu_e}\rangle$ & $\langle \epsilon_{\bar\nu_e}\rangle$ & $\langle \epsilon_{\nu_x}\rangle$ \\
\hline
Spectral M1 & 3.7 & 3.6 & 11.7 & 12.1 & 15.7 & 25.3\\
M1 $\beta \sim 0$ & 3.7 & 3.3 & 11.2 & 11.3 & 13.5 & 26.0 \\
M1 $\beta = 4 $ & 3.6& 3.1 & 11.1 & 12.3& 14.4& 26.4\\
M1 $\beta = 6 $ & 3.6 & 3.1 & 11.0 & 12.5& 14.6& 26.6\\
M1 $\beta = 8 $ & 3.5 & 3.1 & 11.0 & 12.7 & 14.8 & 26.7\\
M1 $\beta \sim \infty$ & 3.2 & 3.1 & 7.4 & 13.5& 15.8 & 34.9\\
Leakage & 13.6 & 5.3 & 9.0 & 11.7 & 14.9 & 22.2\\
\end{tabular}
\end{ruledtabular}
\end{table}

The evolution of the number density itself is a fairly simple process to take into account. The complexities introduced
by uncertainties in the neutrino spectrum and the impact of the choices made in the previous sections on the observed neutrino radiation,
however, make it difficult to estimate how well our scheme will perform in practice. To assess this, we consider a 
test problem for neutrino transport previously used in~\cite{Abdikamalov2012,FoucartM1:2015}. 
We evolve the moments of the neutrino distribution function,
fluid temperature and fluid composition for a 1D profile constructed as a spherical average of a 2D core-collapse simulation $160\,{\rm ms}$
after bounce~\cite{Ott2008}. The velocity of
the fluid is set to zero, and the density profile is assumed to be constant. This test has regions with large scattering opacities
and low absorption opacities for the heavy-lepton neutrinos, and an absorption neutrinosphere close to the scattering neutrinosphere
for the electron neutrinos. The electron antineutrinos lie in between those two extremes. The test thus probe the most problematic
aspects of our algorithm: the evolution of the neutrino average energy in the region in which neutrinos decouple from the fluid.
Due to the lack of velocity and gravitational redshift, we can easily use an energy-dependent neutrino transport scheme
in this problem, giving us a reliable frame of reference to which we can compare our results.

Table~\ref{tab:collapseprof} summarizes our results, listing the neutrino luminosity and average energy $8\,{\rm ms}$ into the simulation.
All simulations evolve the post-bounce profile in octant symmetry, with a low-resolution $50^3$ grid covering a cube of length $300{\rm\,km}$.
We consider an energy-dependent M1 scheme, a leakage scheme, and gray M1 evolutions with various choices of the parameter $\beta$.
As shown in Sec.~\ref{app:flux}, $\beta$ determines how strongly we correct the average energy to account for the fact that low-energy neutrinos
diffuse more easily through high opacity material than high-energy neutrinos. We argued in Sec.~\ref{app:flux} that reasonable values for that
parameter should be $\beta\approx 4-8$. In this test, we find good agreement with the energy-dependent transport scheme for $\beta=4-8$.
The error in the neutrino luminosity remains below $15\%$, and the error in the neutrino average energy remains well below $10\%$.

This is not the case for $\beta\rightarrow\infty$ (practically, we use $\beta = 10^6$ for that simulation). In that case, the average energy of the
heavy-lepton neutrinos is widely overestimated, by nearly $40\%$. Consequently, the heavy lepton neutrinos are more strongly absorbed by the 
fluid and the neutrino luminosity drops by about $40\%$. This is consistent with what we would expect in a situation in which the absorption
neutrinosphere is deeper into the fluid than the scattering neutrinosphere. Without the correction to the neutrino spectrum imposed by $\beta =4-8$, 
the neutrino spectrum is approximated as a blackbody spectrum at the temperature of the absorption neutrinosphere. This completely ignores
the softening of the spectrum due to the much lower diffusion rates of the high-energy neutrinos. More importantly, Table~\ref{tab:collapseprof}  shows
that the results are otherwise very insensitive to the choice of $\beta$. This is highly desirable, as the exact choice of $\beta$ is fairly arbitrary.
Although better than $\beta=10^6$, the choice $\beta=10^{-6}$ leads to larger errors for $\nu_e$ and $\bar\nu_e$ than the favored choices $\beta=4-8$.
Finally, we note that while the leakage scheme provides good energy estimates for $\nu_e$ and $\bar\nu_e$, it otherwise performs much worse than
the M1 schemes. In particular, the $\nu_e$ luminosity is off by a factor of $3.7$.

Another observable in this test is the composition and temperature evolution of the fluid due to neutrino absorption in low-density regions.
Not surprisingly, we find results similar to those for the luminosity and neutrino energy. For $\beta=4-8$, the gray M1 scheme is in good agreement
with the energy-dependent scheme. The choice $\beta = 10^6$ causes excessive neutrino absorption in low-density regions, while the choice 
$\beta=10^{-6}$ underestimate composition changes in the same regions. Overall, this test gives us some confidence that the approximate method
chosen here to estimate neutrino energies can provide reasonable results, and that varying the free parameter $\beta$ in the range $[0-8]$ can 
provide a rough estimate of the uncertainty in the results. Although not by any means a replacement for a truly energy-dependent scheme,
this approximate method can hopefully provide us with better results than the leakage scheme, or the previous iteration of our gray M1 scheme
in which a single average neutrino energy was used everywhere in low opacity regions (see Paper II).

\end{document}